\documentclass[12pt,a4paper]{article}
\usepackage[T1]{fontenc}
\usepackage{lmodern}
\usepackage{amsmath,amssymb,amsfonts}
\usepackage{graphicx,booktabs,float}
\usepackage{caption,subcaption}
\usepackage[a4paper,margin=25mm]{geometry}
\usepackage[section]{placeins}
\usepackage[numbers,sort&compress]{natbib}
\usepackage{hyperref}
\hypersetup{
  colorlinks=true,linkcolor=blue,citecolor=blue,urlcolor=blue,
  pdftitle={Weak-field charged-scalar dynamics in Melvin--Kerr spacetime: an effective Kerr spectral problem},
  pdfauthor={Haryanto M. Siahaan}
}
\newcommand{\be}{\begin{equation}}
\newcommand{\ee}{\end{equation}}

\newcommand{\fdrag}{\varpi}
\newcommand{\mueff}{\mu_{\rm eff}}
\newcommand{\Mw}{M\omega}
\numberwithin{equation}{section}
\numberwithin{figure}{section}

\title{Weak-field charged-scalar dynamics in Melvin--Kerr spacetime:
an effective Kerr spectral problem}
\author{Haryanto M. Siahaan\thanks{\href{mailto:haryanto.siahaan@unpar.ac.id}{\texttt{haryanto.siahaan@unpar.ac.id}}}\\[0.5em]
\small Program Studi Fisika, Universitas Katolik Parahyangan\\
\small Jalan Ciumbuleuit 94, Bandung 40141, Indonesia}
\date{}

\begin{document}
\maketitle

\begin{abstract}
We study the local weak-field reduction of the charged Klein--Gordon
equation in the neutral-seed Melvin--Kerr geometry and the auxiliary
Kerr spectral problem defined by that reduction. At fixed radius and
fixed scalar charge, the metric has no term linear in the magnetic
parameter $b$. In the chosen gauge the first-order Maxwell potential
is $A_\mu=b\,g_{\mu\phi}$, so its interaction with an azimuthal mode
$m$ is absorbed into $\mu_{\rm eff}^{2}=\mu^{2}+2qbm$.
This identity applies to both separated equations and leaves the
gauge-invariant horizon phase in the Kerr form. Extending the
truncated equation to all exterior radii and imposing Kerr outgoing
boundary conditions defines an auxiliary spectrum, evaluated with
Dolan's continued-fraction method. Its linear magnetic response is
the complex mass derivative of the Kerr spectrum. Numerical examples
for the fundamental $\ell=1,2$ branches illustrate this relation,
including the continuation through negative effective mass squared.
The tabulated modes remain damped, and their real-frequency shifts
have opposite signs for the two positive-frequency azimuthal sectors
in the surveyed range. We separate numerical accuracy within the
auxiliary problem from local weak-field power counting and the
unresolved exterior boundary problem. No matching to the exact
magnetized exterior is performed; consequently these frequencies
are not established approximations to global Melvin--Kerr resonances.
\end{abstract}


\section{Introduction}
\label{sec.intro}

Quasinormal modes (QNMs) are the characteristic damped oscillations of
perturbed black holes and provide one of the most direct links between
black-hole perturbation theory and observable ringdown physics. The
foundational stability analyses of Schwarzschild perturbations were given
by Regge and Wheeler \cite{Regge:1957td} and by Zerilli \cite{Zerilli:1970se};
the interpretation of damped oscillations as black-hole normal modes was
developed shortly afterwards \cite{Vishveshwara:1970zz,Press:1971wr}, and
the separability of perturbation equations on Kerr backgrounds was
established through the Teukolsky formalism
\cite{Teukolsky:1973ha,Chandrasekhar:1983}. Modern high-precision QNM
calculations rely on Leaver's continued-fraction method
\cite{Leaver:1985ax}, with extensive subsequent refinements and reviews
\cite{Nollert:1999ji,Kokkotas:1999bd,Berti:2009kk,Konoplya:2011qq}.

Astrophysical black holes are not isolated; they are surrounded by plasma,
accretion flows, jets, and large-scale magnetic fields. A widely used
idealization is Wald's solution, in which an external magnetic field is
treated as a test field on Kerr \cite{Wald:1974np}, an approximation that
underlies the Blandford--Znajek mechanism for jet launching
\cite{Blandford:1977ds}. A geometrically self-consistent alternative is
provided by the Melvin magnetic universe \cite{Melvin:1964} together with
Harrison transformations \cite{Harrison:1968}, which generate exact
Einstein--Maxwell solutions describing black holes immersed in an external
magnetic field. The general magnetization procedure was formulated by Ernst
\cite{Ernst:1976mzr}, and the magnetized Kerr (Melvin--Kerr) solution
was constructed by Ernst and Wild \cite{ErnstWild:1976}; subsequent
work clarified its ergoregions,
near-horizon limits, and thermodynamics
\cite{Gibbons:2013yq,Astorino:2015fna}.

Massive scalar fields around rotating black holes are closely connected
with superradiance and quasi-bound states. The black-hole bomb mechanism
was introduced by Press and Teukolsky \cite{Press:1972zz}, and wave
amplification by rotating black holes was clarified in
\cite{Starobinsky:1973aij}. Massive-field instabilities and quasi-bound
levels were developed in \cite{Zouros:1979iw,Detweiler:1980uk}, and
extended to charged rotating backgrounds in \cite{Furuhashi:2004jk}.
Massive-scalar QNMs and bound states on Schwarzschild and Kerr backgrounds
have been studied with several frequency-domain methods
\cite{Simone:1991wn,Konoplya:2006br,Berti:2005ys,Cardoso:2005vk}. Dolan's
continued-fraction analysis of massive-scalar perturbations of Kerr
\cite{Dolan:2007mj} provides the benchmark spectra used here to validate
the unmagnetized limit.

QNMs and instabilities in magnetized backgrounds have been studied from
several complementary perspectives. For the non-rotating Ernst black hole,
Konoplya and Fontana computed the QNMs of the Ernst black hole in both
time and frequency domains, showing that the external magnetic field
generates an effective scalar mass and modifies the spectrum
\cite{Konoplya:2008b}; the associated magnetically enhanced
superradiant instability of slowly rotating magnetized black holes was
studied in \cite{Konoplya:2007yy}.
Charged massive-scalar QNMs on a Kerr--Newman black hole immersed in an
asymptotically uniform Wald-type magnetic field were computed in
\cite{Kokkotas:2011kn}, where Zeeman splitting and Faraday induction were
identified as distinct spectral effects. Related charged-scalar studies
in magnetized and Ernst-type backgrounds appear in
\cite{Turimov:2019qnm,Wu:2015qqm,Gonzalez:2023Ernst}. Neutral-scalar QNMs of the Ernst--Wild geometry have also been
studied using a perturbative magnetic expansion and a Kerr-like
spectral prescription \cite{Taylor:2025rw}. That calculation concerns
the geometric magnetic correction, whereas the present calculation
isolates the charge coupling linear in $b$. The local charged-scalar
reduction used here was developed in Ref.~\cite{Siahaan:2025plb} in
the bound-state context. Neither the effective-mass idea nor the use
of the massive-scalar Kerr recurrence is new by itself.

The purpose of this paper is to formulate precisely the effective
spectral problem that follows from this reduction, exhibit its gauge
and horizon consistency, and characterize its complex-frequency
response. In the gauge used below, the first-order potential is
$A_\mu=b\,g_{\mu\phi}$. The linear interaction then gives
$(\Box_{\rm Kerr}+\mu^2+2qbm)\Phi=0$ for a mode of azimuthal number
$m$. This is the equilibrium-charge sector of the Wald test field
\cite{Wald:1974np}; at this order there is no gravitational
backreaction of the magnetic field. The exact Harrison construction
specifies the family being expanded, but it does not make the
retained local equation a new global magnetized wave problem.

The auxiliary problem is obtained by extending this equation to
all $r>r_+$ and imposing the massive-scalar Kerr QNM conditions.
Consequently, on a branch continued from Kerr,
$\omega_{\rm aux}(b)=\omega_{\rm Kerr}(\mu^2+2qbm)$, with the other
mode labels held fixed. Its first derivative in $b$ follows from
the Kerr mass derivative; the numerical slope tests check the
implementation of that identity. The numerical examples cover
fundamental co- and counter-rotating branches, direct angular
eigenvalue checks, and continuation through $\mu_{\rm eff}^2=0$.

The physical distinction between the local reduction and this global
extension is essential. The weak-field expansion is not uniform at
large radius, and the exact rotating magnetized exterior is more
complicated than the static Melvin universe
\cite{Gibbons:2013yq}. The continued fraction imposes its outgoing
condition at the formal infinity of the extended Kerr equation; it
does not impose a boundary condition at a finite radius inside a
Kerr-like window. This paper does not determine a global Melvin--Kerr
spectrum or a percentage accuracy for its approximation. Its result
is a defined auxiliary spectral response that can be used in a
future calculation with a physically specified exterior.

The choices $bM\in\{10^{-3},10^{-2}\}$ and $qM=0.1$ are numerical
examples chosen to resolve the linear interaction. They are not
predictions for an astrophysical ringdown signal. The local expansion
requires $|bM|\ll1$, a radial range described below, and, for the
retained charged term to dominate the first omitted metric term
parametrically, $|qM|\gg|bM|$ for nonzero $m$ of order unity.

The paper is organized as follows. Section~\ref{sec.MK} summarizes the
weak-field Melvin--Kerr background, fixes the gauge, and states the
order in which the weak-field and large-radius limits are taken
(Sec.~\ref{subsec.limits}).
Section~\ref{sec.KG} derives the separated charged Klein--Gordon equation.
Section~\ref{sec.QNM} discusses the QNM boundary conditions in the
reference gauge.
Section~\ref{sec.method} describes the continued-fraction implementation
and benchmark validation. Section~\ref{sec.results} presents the
auxiliary spectrum and the effective-potential diagnostics.
Section~\ref{sec.conclusion} concludes. Natural units $c=G_{N}=\hbar=1$
are used throughout.

\section{The weak-field Melvin--Kerr background}
\label{sec.MK}

\subsection{Line element, weak-field truncation, and gauge choice}
\label{subsec.background}

The Melvin--Kerr spacetime describes a rotating black hole immersed in an
external, axisymmetric magnetic field. It is generated from the Kerr seed
by an Ernst--Harrison transformation in Einstein--Maxwell theory and
represents a geometrically self-consistent magnetized black-hole
background, rather than a test electromagnetic field superposed on Kerr.
The exact geometry is not asymptotically flat. For a rotating neutral
seed, it must not be identified globally with the static Melvin
universe: an ergoregion can extend to infinity near the axis
\cite[Secs.~3.2--3.3]{Gibbons:2013yq}. Only a local weak-field
expansion is used here, with $|bM|\ll1$ and coordinates inherited
from Kerr.

In Boyer--Lindquist-type coordinates $(t,r,\theta,\phi)$, the Melvin--Kerr
line element may be written as
\be
ds^{2}
=\frac{\Delta\sin^{2}\!\theta}{f}\,dt^{2}
-\frac{e^{2\gamma}}{f}\!\left(\frac{dr^{2}}{\Delta}+d\theta^{2}\right)
-f\!\left(\fdrag\,dt-d\phi\right)^{2},
\label{MK.metric}
\ee
with $\Delta=r^{2}-2Mr+a^{2}$,
$e^{2\gamma}=\sin^{2}\!\theta\bigl[\Delta a^{2}\cos^{2}\!\theta
+r(r^{3}+a^{2}r+2a^{2}M)\bigr]$, and
$f=e^{2\gamma}/(c_{0}+c_{2}b^{2}+c_{4}b^{4})$,
$c_{0}=r^{2}+a^{2}\cos^{2}\!\theta$. For the neutral Kerr seed the
quadratic coefficient is $c_{2}=2\,{\cal A}\sin^{2}\theta$ with
${\cal A}=(r^{2}+a^{2})^{2}-a^{2}\Delta\sin^{2}\theta$, the
neutral-seed specialization of the magnetized Kerr solution
\cite{Gibbons:2013yq}, and the quartic coefficient $c_{4}$ is given
explicitly in \cite{Siahaan:2025plb}; both enter the metric only at
order $b^{2}$ and beyond, which is beyond the order retained in this
work.
At order $b^{0}$ the geometry is exactly Kerr,
\be
f\big|_{b=0}=\frac{e^{2\gamma}}{r^{2}+a^{2}\cos^{2}\!\theta},
\qquad
\fdrag\big|_{b=0}=\frac{2aMr\sin^{2}\!\theta}{e^{2\gamma}},
\label{f.Kerr.limit}
\ee
with horizons at $r_{\pm}=M\pm\sqrt{M^{2}-a^{2}}$ and Kerr horizon angular
velocity $\Omega_{H}=a/(2Mr_{+})$.

We fix here the dimensions and sign conventions of the external-field
and charge parameters used throughout. In geometric units
($G=c=\hbar=1$), the magnetic-field parameter $b$ of the Harrison
transformation has dimension of inverse length, so that $bM$ is
dimensionless; $b>0$ corresponds to an external field aligned with the
positive rotation axis, i.e.\ parallel to the black-hole spin, and
$b<0$ to the anti-aligned configuration (in the Kerr-like region the
first-order potential has contravariant component $A^{\phi}=b$, the
flat-space potential of a uniform field of strength $B_{0}=2b$ along
the axis; Appendix~\ref{app.horizon}). The scalar charge $q$ enters
through the minimal coupling $(\nabla_{\mu}-iqA_{\mu})$ of
Sec.~\ref{sec.KG} and likewise has dimension of inverse length, with
$qM$ dimensionless; the sign of $q$ is defined relative to this
covariant derivative, so that the physical coupling flips under either
$q\to-q$ or $b\to-b$ and the retained interaction depends on the
parameters only through the product $qb$ (times $m$). In the locally flat Cartesian limit we identify the ordinary
spatial vector potential with $A^i$: $\boldsymbol A=(-by,bx,0)$,
so $\boldsymbol\nabla\times\boldsymbol A=2b\,\hat{\boldsymbol z}$.
This fixes the stated field orientation for the $(+---)$ signature;
$q$ is the signed coupling in the covariant derivative above.
We take $a>0$ throughout, so that $m>0$ labels co-rotating
and $m<0$ counter-rotating modes.

The electromagnetic potential generated by the Harrison transformation has
the stationary, axisymmetric form $A_{\mu}dx^{\mu}=A_{t}\,dt+A_{\phi}\,d\phi$,
with $A_{t}$ and $A_{\phi}$ rational functions of $r$, $\cos\theta$, $M$,
$a$, and $b$, given explicitly in \cite{Siahaan:2025plb}. Two features of
this gauge potential are central to the present work. First, since the Kerr seed is electrically neutral, the gauge field is
linear in $b$ at leading order:
$A_{t}=b\,{\cal A}_{t}^{(1)}(r,\theta)+\mathcal{O}(b^{3})$ and
$A_{\phi}=b\,{\cal A}_{\phi}^{(1)}(r,\theta)+\mathcal{O}(b^{3})$. The
interaction of a charged scalar field with the background gauge field
therefore first appears at order $bq$. To yield the truncation fully
explicit, the weak-field scheme used throughout this paper
\emph{retains} (i) the exact Kerr metric, i.e.\ the order-$b^{0}$ part
of \eqref{MK.metric}, and (ii) the $\mathcal{O}(bq)$ interaction terms
linear in $qA_{\mu}$ in the charged Klein--Gordon equation; it
\emph{discards} (iii) the $\mathcal{O}(b^{2})$ corrections to the
metric functions (the terms $c_{2}b^{2}$ and $c_{4}b^{4}$ in $f$, and
the corresponding corrections to $\fdrag$ and hence to the metric
components), (iv) the $\mathcal{O}(b^{3}q)$ terms linear in the scalar
charge, generated by the interference of the $\mathcal{O}(b^{2})$
metric corrections with the $\mathcal{O}(bq)$ interaction and by the
$\mathcal{O}(b^{3})$ part of the potential, and (v) the
$\mathcal{O}(b^{2}q^{2})$ contact term $q^{2}A_{\mu}A^{\mu}$. No
$\mathcal{O}(b^{2}q)$ sector arises: for the neutral Kerr seed the
metric is even and the potential odd under $b\to-b$
(cf.\ Sec.~\ref{subsec.precision}).

Second, a residual constant in $A_t$ must be fixed when comparing
coordinate frequencies. At first order the Harrison gauge of
Ref.~\cite{Siahaan:2025plb} differs from our gauge by
\be
A_t^{\rm(Harrison)}=A_t+\frac{4M^2b}{a}+\mathcal O(b^3).
\label{At.Harrison.inf}
\ee
For $a\ne0$ this is removed by
\be
A_t^{\rm(Harrison)}\longrightarrow A_t^{\rm(Harrison)}-\frac{4M^2b}{a}.
\label{gauge.shift}
\ee
The resulting first-order potential is smooth as $a\to0$ and obeys
$A_t^{(b)}\to0$ at the formal large-radius end of the Kerr expansion.
We call this the \emph{reference gauge}. This large-radius statement
concerns the coefficient of $b$ at fixed coordinates, subsequently
extended as a Kerr field. It is not a statement about the exact
fixed-$b$ asymptotic potential. The first-order azimuthal component
$A_\phi^{(b)}\sim-br^2\sin^2\theta$ grows with radius, although
$A^{(b)\phi}=b$ is constant. Its linear interaction is therefore
finite, while the discarded quadratic interaction grows as
$q^2b^2r^2\sin^2\theta$ in this local expansion.

Because the scalar field under consideration is charged, the mode frequency $\omega$ is
gauge \emph{covariant} rather than gauge invariant under the residual
freedom of constant shifts of $A_{t}$: under $A_{t}\to A_{t}+C$ (with
$C$ real constant), accompanied by the field redefinition
$\Phi\to e^{iqCt}\Phi$, the frequency of any given mode transforms as
\be
\omega\;\longrightarrow\;\omega-qC,
\label{gauge.dictionary}
\ee
while $\Im\,\omega$ is unchanged. Therefore, it is useful to
distinguish the gauge-dependent intermediate quantities from the
gauge-invariant physical statements. Gauge invariant are: (i) the
imaginary part $\Im\,\omega$ of every mode, and hence the damping or growth rate of a specified mode; (ii) the combination
$\omega+qA_{t}(r\to\infty)$; and (iii) for fixed scalar charge $q$,
differences of frequencies on the same background with the same constant gauge shift. For differences between backgrounds at different $b$, the reference gauge must be fixed consistently; a $b$-dependent gauge constant would change the real-frequency difference. All frequencies quoted in this paper
refer to the gauge \eqref{gauge.shift}, in which $A_{t}(\infty)=0$, so
that the quoted $\omega$ coincides with the invariant combination
$\omega+qA_{t}(\infty)$ and connects smoothly to the standard Kerr QNM
frequency as $b\to 0$ and to the Schwarzschild one as $a\to 0$. To
compare with a computation performed in a different gauge one applies
the dictionary \eqref{gauge.dictionary}; for example, in the unshifted
Harrison gauge \eqref{At.Harrison.inf} the same physical modes appear
at $\omega^{\rm (Harrison)}=\omega-4qM^{2}b/a$, the difference being
precisely the pure-gauge, spin-singular constant removed by
\eqref{gauge.shift}, which carries no dynamical information.

Since the spacetime possesses no asymptotically flat region, it is
worth stating the operational meaning of $\omega$ precisely. The
frequency $\omega$ in the ansatz \eqref{ansatz} is defined as the
Fourier conjugate of the Boyer--Lindquist-type time coordinate $t$ of
the line element \eqref{MK.metric}, in the gauge \eqref{gauge.shift}
with $A_{t}(\infty)=0$. Four related notions should be
distinguished. (i) The \emph{coordinate frequency} in an arbitrary
residual gauge, which shifts according to the dictionary
\eqref{gauge.dictionary} and has no invariant meaning by itself.
(ii) The \emph{gauge-invariant combination}
$\omega+qA_{t}(r\to\infty)$, which coincides with the quoted $\omega$
in the gauge adopted here. (iii) The frequency that would be
\emph{measured} locally by a stationary observer with four-velocity
$u^{\mu}=u^{t}[(\partial_{t})^{\mu}+\Omega_{\rm obs}(\partial_{\phi})^{\mu}]$,
which is the gauge-invariant local phase frequency
\be
\omega_{\rm loc}=u^{t}\Bigl[\,\omega-m\,\Omega_{\rm obs}
+q\bigl(A_{t}+\Omega_{\rm obs}A_{\phi}\bigr)\Bigr].
\label{omega.loc}
\ee
A local observer therefore does not measure $\omega$ itself; only for
a sufficiently distant, approximately static observer in the
Kerr-like window, with $u^{t}\to 1$, $\Omega_{\rm obs}\to 0$, and
$qA_{t}\to 0$ in the present gauge, does $\omega_{\rm loc}$ approach
the quoted $\omega$. (iv) A hypothetical frequency defined at the
exact magnetized exterior, which does not correspond to any observer
used here: no such global asymptotic notion is invoked in this paper.
In all occurrences of $A_t(\infty)$ in the frequency dictionary,
the limit is that of the first-order Kerr potential, not of the
exact solution at fixed $b$.
All quoted values are coordinate frequencies in the fixed gauge
\eqref{gauge.shift} with $A_{t}(\infty)=0$; in this gauge they
coincide numerically with the gauge-invariant combination of item
(ii). Locally observed frequencies follow from \eqref{omega.loc}.

\subsection{Order of limits: weak-field expansion versus the
large-radius limit}
\label{subsec.limits}

Because the exact Melvin--Kerr spacetime is not asymptotically flat,
two operations enter the formulation of any spectral problem on
it---the weak-field expansion in $b$ and the large-radius limit
$r\to\infty$---and they do not commute: the weak-field expansion is
not uniform at large radius. The order in which they are performed
therefore \emph{defines} the spectral problem, and we state it here
explicitly, since every statement about boundary conditions in this
paper rests on it.

\emph{Large radius at fixed field.} For any fixed $b\neq 0$ the
geometry \eqref{MK.metric} is not asymptotically Kerr. Its rotating neutral-seed exterior is not globally the static Melvin universe \cite{Gibbons:2013yq}.
Writing the magnetized function as $f=f_{\rm Kerr}/|\Lambda|^{2}$,
with $f_{\rm Kerr}=e^{2\gamma}/\Sigma$, $\Sigma\equiv
r^{2}+a^{2}\cos^{2}\theta$, and
$|\Lambda|^{2}=(c_{0}+c_{2}b^{2}+c_{4}b^{4})/c_{0}$, the coefficient of
the first $dt^{2}$ term in \eqref{MK.metric} (the full component being
$g_{tt}=\Delta\sin^{2}\theta/f-f\fdrag^{2}$) reads, at fixed angle
away from the symmetry axis,
\be
\frac{\Delta\sin^{2}\theta}{f}
=\frac{\Delta\,\Sigma\,\sin^{2}\theta}{e^{2\gamma}}\;|\Lambda|^{2},
\qquad
|\Lambda|^{2}\;\xrightarrow[\;r\gg b^{-1}\;]{}\;b^{4}r^{4}\sin^{4}\theta ,
\label{gtt.Melvin}
\ee
the prefactor tending to unity while $|\Lambda|^{2}$ exhibits the
quartic Melvin growth: the explicit coefficients of
\cite{Siahaan:2025plb} give $c_{4}\sim r^{6}\sin^{4}\theta$ and
$e^{2\gamma}\sim r^{4}\sin^{2}\theta$ at large $r$, the quartic
coefficient $b^{4}r^{4}\sin^{4}\theta=(B_{0}\,r\sin\theta/2)^{4}$
being that of the Melvin universe with asymptotic field strength
$B_{0}=2b$ (Appendix~\ref{app.horizon}). Hence
\be
\lim_{r\to\infty}\,\lim_{b\to 0}\;\frac{\Delta\sin^{2}\theta}{f}=1,
\qquad\text{whereas}\qquad
\lim_{r\to\infty}\;\frac{\Delta\sin^{2}\theta}{f}\,\Big|_{b\neq 0}
=\infty ,
\label{noncommuting}
\ee
and no subsequent $b\to 0$ renders the second limit finite. The same
non-commutativity is present in the matter sector: the retained
interaction $2qbm$ is a constant, whereas the omitted contact term
$-q^{2}A_{\mu}A^{\mu}\simeq q^{2}b^{2}r^{2}\sin^{2}\theta$ grows
without bound at fixed $b$. These observations demonstrate nonuniformity of the inner expansion;
they do not determine the exact outer wave operator. In particular,
Eq.~\eqref{gtt.Melvin} concerns a metric factor at fixed angle away
from the axis, not the full $g_{tt}$ or the near-axis asymptotics.
The static Melvin flux-tube picture is insufficient to prescribe
the outer conditions for the rotating neutral-seed solution.
Those conditions require a separate analysis.

\emph{Weak field at fixed radius.} Conversely, at any fixed
$r/M$---including the horizon and the photon-sphere region---the
Harrison-generated metric and potential are regular power series in
$b$,
\be
g_{\mu\nu}=g^{(0)}_{\mu\nu}+b^{2}g^{(2)}_{\mu\nu}+\mathcal{O}(b^{4}),
\qquad
A_{\mu}=b\,{\cal A}^{(1)}_{\mu}+\mathcal{O}(b^{3}),
\label{weakfield.series}
\ee
with $g^{(0)}_{\mu\nu}$ exactly Kerr and
$b^{2}g^{(2)}_{\mu\nu}/g^{(0)}_{\mu\nu}=\mathcal{O}(b^{2}r^{2})$. The
$\mathcal{O}(bq)$ equation of Sec.~\ref{sec.KG} is the charged
Klein--Gordon operator expanded to first order in $b$. The coordinates and
$a/M$, $qM$, $\mu M$ are held fixed; the first-order term is entirely the
charged interaction $\propto qb$, since the metric has no
$\mathcal{O}(b)$ part. It is a local inner weak-field equation: its neglected terms are
small uniformly on any fixed range of $r/M$ as $bM\to 0$, the
relevant local ratios scaling as $(br)^{2}$ (metric sector) and
$|qb|r^{2}/|m|$ (contact term relative to the retained interaction,
for $m\neq 0$). A sufficient parametric radial window for retaining the linear interaction ahead of these omitted terms is bounded by the smaller of the two scales,
\be
M\lesssim r\ll L_{\rm out},\qquad
L_{\rm out}\equiv\min\!\left(|b|^{-1},\
\sqrt{\frac{|m|}{|qb|}}\right),
\label{intermediate.region}
\ee
where the first restriction comes from the $\mathcal{O}(b^{2})$ metric
corrections $\sim b^{2}r^{2}$ and the second from the omitted contact
term $q^{2}A_{\mu}A^{\mu}\sim q^{2}b^{2}r^{2}$ measured against the
retained linear term $\sim|qbm|$. For the survey parameters $qM=0.1$,
$bM=10^{-2}$, $|m|=1$ the local breakdown scale is
$L_{\rm out}\simeq 31.6M$, compared with photon-sphere radii of order
$3M$ and the purely geometric scale $b^{-1}=100M$; the asymptotic
condition $r\ll L_{\rm out}$ should not be read as validity all the
way to $31.6M$.

\emph{Order adopted in this paper.} Let $\mathcal{L}(b)$ denote the
charged Klein--Gordon operator, with the coordinates and all other
parameters held fixed. The spectral problem solved below is defined
in two steps. (i) The operator is truncated at first order in $b$ at
fixed $r/M$,
\be
\mathcal{L}_{\rm aux}(b)=\mathcal{L}(0)
+b\,\frac{\partial\mathcal{L}}{\partial b}\Big|_{b=0},
\qquad 0<|bM|\ll 1,
\label{order.of.limits}
\ee
with $b$ small but finite (a literal $b\to 0$ would remove the
retained $2qbm$ term); this yields the inner equation, which is the
Kerr massive-scalar equation with $\mu^{2}\to\mueff^{2}$
(Sec.~\ref{sec.KG}). (ii) That equation is \emph{analytically
extended} to all $r>r_{+}$, and the standard Kerr QNM boundary
conditions---ingoing at the horizon and outgoing at infinity---are
imposed at its own endpoints (Sec.~\ref{sec.QNM}). In the language
of limits: the large-radius limit is taken of the $b$-expanded
operator, never of the exact one. Equation \eqref{noncommuting}
makes the consequence explicit: after step (i) the large-radius
limit is the Kerr one, so the extended equation is asymptotically
flat by construction and its effective potential tends to the
constant $\omega^{2}-\mueff^{2}$ (Sec.~\ref{sec.KG}). That constant
is the signature of the prescription \eqref{order.of.limits}---Kerr
asymptotics with a shifted mass-squared parameter---and is not the
large-radius behaviour of the Melvin--Kerr spacetime, for which no
such constant exists. The reverse order, (ii) before (i), would
require solving the exact far-region problem with its own
boundary conditions and would define a different spectrum.
Equivalently, the construction adopted here does not interchange two
limits of one and the same exact boundary-value problem: it defines
a different global problem after the local expansion, whose outer
condition is imposed at the formal infinity of the extended equation
and not at a finite radius inside the inner validity window
\eqref{intermediate.region}.

\emph{Status of the computed frequencies.} This prescription defines
the QNMs of an auxiliary effective Kerr problem. No outer matching
or independently specified finite-radius exterior is solved here.
Local agreement of differential operators does not establish
agreement of their spectra when the outer conditions differ.
The auxiliary frequencies therefore have no demonstrated error
bound relative to global Melvin--Kerr resonances. All spectral
precision statements below refer to the auxiliary problem.

\section{Separation of the Klein--Gordon equation}
\label{sec.KG}

A massive charged scalar field $\Phi$ on the weak-field Melvin--Kerr
background obeys the minimally coupled Klein--Gordon equation
\be
\bigl[(\nabla_{\mu}-iqA_{\mu})(\nabla^{\mu}-iqA^{\mu})+\mu^{2}\bigr]\Phi=0,
\label{KG.eq}
\ee
where $\mu$ and $q$ are the scalar mass and charge; the sign of the
mass term is that of the signature $(+---)$ of \eqref{MK.metric}, for
which the flat-space limit gives $\omega^{2}=\boldsymbol{k}^{2}+\mu^{2}$.
The local reduction can be seen before separating variables. In the
reference gauge the first-order potential satisfies
\be
A^{(b)\mu}=b\,\delta^\mu_\phi,\qquad
\nabla_\mu A^{(b)\mu}=0.
\label{axial.potential.main}
\ee
The second identity follows from axisymmetry, or from the Killing
equation. The linear interaction in the wave operator is
$-iq(2A^\mu\partial_\mu+\nabla_\mu A^\mu)$, so on an azimuthal mode
$e^{im\phi}$ it equals $+2qbm$. Thus, within the local first-order
truncation,
\be
\bigl(\Box_{\rm Kerr}+\mu^2+2qbm\bigr)\Phi=0.
\label{local.reduction}
\ee
This statement relies on the chosen charge sector as well as the
reference gauge; a generic Wald field with an independent black-hole
charge need not have $A^{(b)t}=0$. Appendix~\ref{app.horizon}
gives the covariant components and the horizon check.

With the separated
ansatz
\be
\Phi(t,r,\theta,\phi)=e^{i(m\phi-\omega t)}\,R_{\ell m}(r)\,S_{\ell m}(\theta),
\label{ansatz}
\ee
and using the reference gauge \eqref{gauge.shift}, the
weak-field truncation of \eqref{KG.eq} (keeping all $b^{0}$ and $bq$
contributions and discarding terms of order $b^{2}$, $b^{2}q^{2}$, and
$b^{3}q$; no $b^{2}q$ terms arise, by the parity noted in
Sec.~\ref{sec.MK}) reduces, after multiplying through by an overall factor of
$-\Sigma=-(r^{2}+a^{2}\cos^{2}\theta)$, to
\begin{equation}
\begin{aligned}
&\frac{1}{R_{\ell m}}\frac{d}{dr}\!\left(\Delta\frac{dR_{\ell m}}{dr}\right)
+\frac{1}{\sin\theta\,S_{\ell m}}\frac{d}{d\theta}\!\left(\sin\theta
\frac{dS_{\ell m}}{d\theta}\right)\\
\quad{}&+\bigl[f_{\mu^{2}}\mu^{2}+f_{\omega^{2}}\omega^{2}+f_{\omega}\omega
+f_{m^{2}}m^{2}+f_{m}m+f_{0}\bigr]=0,
\end{aligned}
\label{KG.general.structure}
\ee
where the coefficient functions $f_{(*)}$ in the reference gauge take the form \cite{Siahaan:2025plb} (re-derived from the
first-order potential in Appendix~\ref{app.horizon})
\begin{align}
f_{\mu^{2}}&=-r^{2}-a^{2}\cos^{2}\theta,
\label{fmu.PLB}\\
f_{\omega^{2}}&=a^{2}\cos^{2}\theta
+\frac{r(2a^{2}M+a^{2}r+r^{3})}{\Delta},
\label{fomega2.PLB}\\
f_{\omega}&=-\frac{4aMmr}{\Delta},
\label{fomega.PLB}\\
f_{m^{2}}&=\frac{a^{2}}{\Delta}-\frac{1}{\sin^{2}\theta},
\label{fm2.PLB}\\
f_{m}&=-2bq\,(r^{2}+a^{2}\cos^{2}\theta),
\label{fm.PLB}\\
f_{0}&=0.
\label{f0.PLB}
\end{align}
The crucial feature of these coefficients is that the magnetic correction
appears \emph{exclusively} in $f_{m}$, with the simple structure
$-2bq\,c_{0}$, where $c_{0}=r^{2}+a^{2}\cos^{2}\theta$ is the
Boyer--Lindquist $\Sigma$ function. There is no $\omega$-dependent magnetic
correction in $f_{\omega}$, no $1/\Delta$ pole in $f_{m}$, and no $1/a$
scaling. These three structural simplifications are direct consequences of
the gauge choice \eqref{gauge.shift}.

Before separating variables, it is worth exhibiting the algebraic
origin of the effective-mass structure. Comparing \eqref{fmu.PLB} and
\eqref{fm.PLB}, the entire $\mathcal{O}(bq)$ interaction multiplies
\emph{precisely the same function} of $(r,\theta)$ as the mass term:
\be
f_{\mu^{2}}\,\mu^{2}+f_{m}\,m
=-(r^{2}+a^{2}\cos^{2}\theta)\bigl(\mu^{2}+2qbm\bigr)
=f_{\mu^{2}}\,\bigl(\mu^{2}+2qbm\bigr).
\label{fm.proportionality}
\ee
Already at the level of \eqref{KG.general.structure}, therefore, the
weak-field magnetized equation \emph{is} the Kerr massive-scalar
equation with the replacement $\mu^{2}\to\mu^{2}+2qbm$, and the
separation of variables must proceed exactly as in the Kerr
massive-scalar problem. Furthermore, it is instructive to track
explicitly how the two pieces of
$f_{m}m=-2qbm\,(r^{2}+a^{2}\cos^{2}\theta)$ are distributed between
the two separated equations. The $\theta$-dependent piece,
$-2qbm\,a^{2}\cos^{2}\theta$, joins the mass and frequency terms of
the angular sector, where it completes the spheroidicity combination,
\be
a^{2}(\omega^{2}-\mu^{2})\cos^{2}\theta-2qbm\,a^{2}\cos^{2}\theta
=a^{2}\bigl(\omega^{2}-\mueff^{2}\bigr)\cos^{2}\theta ,
\label{angular.distribution}
\ee
while the $r$-dependent piece, $-2qbm\,r^{2}$, joins the radial mass
term, which it completes \emph{exactly},
\be
-\mu^{2}r^{2}-2qbm\,r^{2}
=-\mueff^{2}\,r^{2},
\label{radial.distribution}
\ee
with no leftover term of any kind. Since \emph{every} term generated
by the magnetic interaction is accounted for by $\mueff^{2}$ in
\eqref{angular.distribution}--\eqref{radial.distribution}, no
finite-radius ($r$-dependent) magnetic residue---and, in the grouping
of \eqref{radial.eq} below, not even a constant one---remains in the
radial equation at $\mathcal{O}(bq)$.

The separated radial equation reads
\be
\frac{d}{dr}\!\left(\Delta\frac{dR_{\ell m}}{dr}\right)
+\left[\frac{H^{2}}{\Delta}+2ma\omega-a^{2}\omega^{2}-\mu^{2}r^{2}
-2qbm\,r^{2}-K_{\ell m}\right]R_{\ell m}=0,
\label{radial.eq}
\ee
with $H=(r^{2}+a^{2})\omega-am$ and $K_{\ell m}$ the separation
constant; the grouping of the constant terms in \eqref{radial.eq} is
that of Dolan \cite{Dolan:2007mj}, in which $K_{\ell m}$ coincides
with the spheroidal eigenvalue of the angular equation below.
Introducing
\be
\mueff^{2}\equiv\mu^{2}+2qbm
\label{mueff.def}
\ee
turns \eqref{radial.eq} into \emph{exactly} the Kerr massive-scalar
radial equation of \cite{Dolan:2007mj} with $\mu^{2}\to\mueff^{2}$, with the
same definition of the separation constant, evaluated at the new spheroidicity:
\be
\frac{d}{dr}\!\left(\Delta\frac{dR_{\ell m}}{dr}\right)
+\left[\frac{H^{2}}{\Delta}+2ma\omega-a^{2}\omega^{2}-\mueff^{2}r^{2}
-K_{\ell m}\right]R_{\ell m}=0.
\label{radial.eq.muEff}
\ee
The same mass-squared parameter $\mueff^{2}$ enters the angular sector. The
separated angular equation
\be
\frac{1}{\sin\theta}\frac{d}{d\theta}\!\left(\sin\theta
\frac{dS_{\ell m}}{d\theta}\right)
+\left[K_{\ell m}+a^{2}(\omega^{2}-\mueff^{2})\cos^{2}\theta
-\frac{m^{2}}{\sin^{2}\theta}\right]S_{\ell m}=0,
\label{angular.theta}
\ee
takes, with $x=\cos\theta$, the standard generalized spheroidal form
\be
\frac{d}{dx}\!\left[(1-x^{2})\frac{dS_{\ell m}}{dx}\right]
+\left[\Lambda_{\ell m}+c^{2}\,x^{2}-\frac{m^{2}}{1-x^{2}}\right]S_{\ell m}=0,
\label{spheroidal.eq}
\ee
with
\be
\Lambda_{\ell m}=K_{\ell m},\qquad
c^{2}=a^{2}(\omega^{2}-\mueff^{2}).
\label{Lambda.c2.def}
\ee
Equation~\eqref{Lambda.c2.def} fixes our convention, matching that of
\cite{Dolan:2007mj}: for real $\omega$ in the massless limit one has
$c^{2}=a^{2}\omega^{2}>0$; for QNMs, $\omega$ and hence $c^{2}$ are
complex, and the angular eigenvalue is defined by analytic
continuation in this convention.

The structural conclusion of this separation is that the magnetized
weak-field charged-scalar problem on Melvin--Kerr, in the reference gauge, is mathematically equivalent to the Kerr massive-scalar
problem under the single substitution $\mu^{2}$ to $\mueff^{2}$ introduced in (\ref{mueff.def}) applied uniformly to the radial endpoint exponent and to the angular
spheroidicity parameter. The value of $K_{\ell m}$ changes with $c^2$; only its convention and functional eigenvalue problem are unchanged. No
magnetic residue of any kind---$r$-dependent or constant---remains in
the radial equation \eqref{radial.eq.muEff}, as guaranteed by the
proportionality \eqref{fm.proportionality} and the explicit
distribution
\eqref{angular.distribution}--\eqref{radial.distribution}. For
small $|c^{2}|$, the angular eigenvalue admits the perturbative
expansion (Seidel \cite{Seidel:1989bp}, Berti--Cardoso--Casals
\cite{Berti:2005gp})
\be
\Lambda_{\ell m}\simeq \ell(\ell+1)
+\frac{2c^{2}\bigl[m^{2}+\tfrac{1}{2}-\ell(\ell+1)\bigr]}{(2\ell+3)(2\ell-1)},
\label{lambda.approx}
\ee
which we use throughout the numerical survey. For the parameters considered
($qM\le 0.1$, $bM\le 10^{-2}$, $|\Mw|\sim 0.3$), $|c^{2}|\lesssim 0.1$,
and the next-order term in \eqref{lambda.approx} is of order $10^{-4}$ in
$\Lambda_{\ell m}$. The corresponding shift in the QNM frequency is below
$10^{-5}$ in $\Mw$ and below the precision retained in the spectral
tables.

For diagnostic purposes we also record the effective-potential form of
the radial equation. With the tortoise coordinate
$dy/dr=(r^{2}+a^{2})/\Delta$ and the rescaled radial function
$Y_{\ell m}=\sqrt{r^{2}+a^{2}}\,R_{\ell m}$,
\be
\frac{d^{2}Y_{\ell m}}{dy^{2}}+V_{\rm eff}(r)\,Y_{\ell m}=0,
\label{Schrodinger.form}
\ee
\be
V_{\rm eff}(r)=\frac{\Delta}{(r^{2}+a^{2})^{4}}
\Bigl[(r^{2}+a^{2})^{2}\,{\cal K}(r)
-a^{2}(r^{2}+a^{2})+2rM(2a^{2}-r^{2})\Bigr],
\label{Veff}
\ee
\be
{\cal K}(r)=\frac{H^{2}}{\Delta}+2ma\omega-a^{2}\omega^{2}
-\mu^{2}r^{2}-2qbm\,r^{2}-K_{\ell m}.
\label{Kcal.def}
\ee
At fixed reference frequency, the magnetic dependence enters through both
$-2qbm\,r^2$ and $K_{\ell m}(c^2)$. Their combined contribution is
included in the potential diagnostics below.
At large radius of the analytically extended equation,
$V_{\rm eff}\to\omega^{2}-\mueff^{2}$, regular at $a=0$. This
constant is the $r\to\infty$ limit taken \emph{after} the weak-field
expansion, following the prescription \eqref{order.of.limits}: it is the Kerr
large-radius behaviour with a shifted mass-squared parameter, not the
large-radius behaviour of the exact Melvin--Kerr wave operator, whose
coefficients grow without bound (Sec.~\ref{subsec.limits}).

\section{Quasinormal-mode boundary conditions}
\label{sec.QNM}

We adopt the convention $\omega=\omega_{R}+i\omega_{I}$ with the time
dependence $e^{-i\omega t}$, so that $\omega_{I}<0$ corresponds to a damped
mode. This is the convention used in Dolan's tabulated spectra
\cite{Dolan:2007mj}. Boundary conditions are imposed on the
analytically extended auxiliary equation \eqref{Schrodinger.form}, as
defined in Sec.~\ref{subsec.limits}. Its tortoise coordinate maps
$r=r_{+}$ to $y\to-\infty$ and the formal outer endpoint $r\to\infty$
to $y\to+\infty$; the latter is not an asymptotic region in which the
truncated equation approximates the full geometry.

At the horizon, the appropriate boundary condition is the purely ingoing
wave condition
\be
Y_{\ell m}\sim e^{-ik_{+}y},\qquad y\to-\infty,
\label{QNM.hor}
\ee
where
\be
k_{+}\equiv\omega-m\Omega_{H}+q\,\chi^{\mu}A_{\mu}\big|_{r_{+}},
\label{kplus.def}
\ee
which reduces to $k_{+}=\omega-m\Omega_{H}$ at $\mathcal{O}(bq)$ in
the reference gauge, as demonstrated explicitly in
Appendix~\ref{app.horizon}, and where the Leaver-variable horizon
exponent
$\sigma=2Mr_{+}(\omega-\omega_{c})/(r_{+}-r_{-})$, with
$\omega_{c}=m\Omega_{H}=ma/(2Mr_{+})$ (in the units $M=1$ used by
Dolan this reduces to his expression
$\sigma=2r_{+}(\omega-\omega_{c})/(r_{+}-r_{-})$), is the same as for
the unmagnetized Kerr massive-scalar problem
\cite{Dolan:2007mj,Leaver:1985ax}. This follows directly from
\eqref{radial.eq.muEff}: at the order retained, the radial
equation is structurally identical to the Kerr massive-scalar radial
equation under the parameter substitution $\mu^{2}\to\mueff^{2}$, and its
near-horizon behavior is therefore controlled by the same $\sigma$.
The mass-squared parameter $\mueff^{2}$ does not enter $\sigma$ since it appears in the
radial equation only through terms regular at $r=r_{+}$.

We do not impose an additional finite-radius magnetic correction at
the horizon. Within the projected $\mathcal{O}(bq)$ separable radial
equation \eqref{radial.eq.muEff} used here, the near-horizon indicial
equation contains no additional $bq/\Delta$ pole: the magnetic terms
in the separated radial coefficients are regular at $r=r_{+}$.
The QNM horizon exponent is therefore the same as in the unmagnetized
massive-scalar Kerr problem, while the retained magnetic dependence
enters through the parameter substitution $\mu^{2}\to\mueff^{2}=
\mu^{2}+2qbm$ in the asymptotic exponent and in the angular
spheroidicity. This is the unique $\mathcal{O}(bq)$ deformation
consistent with the separable form of the radial equation in the
reference gauge.

The absence of a magnetic correction to the horizon exponent, in the
frequency of the adopted gauge, follows from the gauge-invariant
near-horizon combination demonstrated in
Appendix~\ref{app.horizon}; the horizon Killing potential by itself
changes under a constant shift of $A_{t}$. There we display the
$\mathcal{O}(b)$ gauge potential in closed form, show that in the
reference gauge the horizon Killing potential
$\chi^{\mu}A_{\mu}|_{r_{+}}$, with
$\chi=\partial_{t}+\Omega_{H}\partial_{\phi}$ the horizon generator,
vanishes identically at $\mathcal{O}(b)$, and exhibit how the
gauge-invariant near-horizon combination
$\omega-m\Omega_{H}+q\,\chi^{\mu}A_{\mu}|_{r_{+}}$ reduces to the bare
Kerr form $\omega-m\Omega_{H}$ when expressed through the frequency of
the reference gauge. The same appendix traces the physical
origin of this cancellation to an exact interplay between the charge
induced by the Harrison transformation and the residual gauge constant,
and verifies consistency with the Harrison-gauge horizon momentum of
the companion analysis \cite{Siahaan:2025plb}.

At the large-radius end of the analytically extended inner equation
[after the first-order truncation \eqref{order.of.limits}], in the
same gauge,
\be
k_{\infty}^{2}=\omega^{2}-\mueff^{2}=\omega^{2}-\mu^{2}-2qbm,
\label{kinfty}
\ee
which is regular at $a\to 0$. Equation~\eqref{kinfty} is the
dispersion relation of the truncated effective-Kerr equation; it is
not obtained by taking $r\to\infty$ at fixed $b$ in the exact
Melvin--Kerr equation and does not imply asymptotic flatness of that
geometry. The outer QNM condition selects the outgoing exponential
\be
Y_{\ell m}\sim e^{+ik_{\infty}y},\qquad y\to+\infty,
\label{QNM.inf}
\ee
which specifies the leading exponential factor; the accompanying
massive-Kerr power-law factors are retained in the Leaver ansatz
\eqref{series.ansatz}. For complex $\omega$, $k_{\infty}$ is
generally complex; the outgoing branch is defined by analytic
continuation from the massless or unmagnetized Kerr QNM branch.

The sign of $\mueff^{2}-\mu^{2}=2qbm$ is positive for co-rotating modes
($m=+1$) with $qb>0$ and negative for counter-rotating modes ($m=-1$).
For $\mu M=0$ and the parameters surveyed below,
$\mueff^{2}M^{2}=2qbm M^{2}$ is positive of order $10^{-3}$ for $m=+1$ and
negative of the same magnitude for $m=-1$. The negative-$\mueff^{2}$ case
is allowed for QNMs because $k_{\infty}=\sqrt{\omega^{2}-\mueff^{2}}$ is
well-defined for complex $\omega$ regardless of the sign of $\mueff^{2}$;
this is unlike the bound-state regime, where $\omega^{2}<\mueff^{2}$ on
the real axis is required and $\mueff^{2}>0$ enters as a positivity
constraint \cite{Siahaan:2025plb}.

A negative $\mueff^2$ is an effective parameter of this extended
equation, not a negative physical rest mass in the original
charged-scalar theory. The tracked Kerr QNM branch can be continued
through $\mueff^2=0$ without encountering the square-root threshold
$\omega^2=\mueff^2$ at the parameters studied. This is a statement
about a particular resonance branch. It does not establish stability
of the complete auxiliary initial-value problem: extending a
negative mass-squared term to infinity can change its low-frequency
sector, which is not surveyed here. Nor does it establish stability
or instability of the exact magnetized spacetime. That question
requires the omitted terms and the physical exterior conditions.
Every tabulated continuation has $\Im\omega<0$, so those particular
modes remain damped.

The sign-flip of $\mueff^{2}-\mu^{2}$
between the two rotating sectors is the leading driver of the
opposite-sign pattern in the spectral shifts seen in
Sec.~\ref{sec.results}.

The robustness of the analytic continuation defining the outgoing
branch for $\mueff^{2}<0$ can be tested directly, by following the
fundamental root continuously as the product $qbm$ passes through
zero. We note first that the crossing of $\mueff^{2}=0$ is a
\emph{sign crossing} of an equation parameter, not a branch point of
the continuation: the square-root branch point of
$\varrho=\sqrt{\mueff^{2}-\omega^{2}}$ sits at $\mueff^{2}=\omega^{2}$,
which for $|\Mw|\sim 0.27$ and $|\mueff^{2}M^{2}|\le 2\times 10^{-3}$
is never approached. Table~\ref{tab.branch} reports the sweep for the
counter-rotating massless mode at $a=0.3M$, $qM=0.1$, with $bM$ varied
from $-10^{-2}$ to $+10^{-2}$, so that $\mueff^{2}M^{2}$ runs from
$+2\times 10^{-3}$ through zero to $-2\times 10^{-3}$. The root moves
smoothly and approximately linearly through $\mueff^{2}=0$, with no jump, kink, or
loss of convergence at the sign crossing. We also verified
implementation independence of the branch choice: re-solving each row
with the two natural prescriptions for the sign of
$\varrho=\sqrt{\mueff^{2}-\omega^{2}}$---(a) the principal square root
followed by the half-plane selection rule of
Sec.~\ref{sec.method}, and (b) continuity tracking of $\varrho$ along the
sweep---returns identical frequencies at all displayed digits at every
point (the two prescriptions are implemented independently in our
numerics). The continuation is therefore
unambiguous and stable throughout the surveyed range, as expected from
the parametric smallness
$|2qbm|M^{2}\le 2\times 10^{-3}\ll|\omega^{2}|M^{2}$, which keeps the
roots far from the actual branch point of $\varrho$.

\begin{table}[htbp]
\centering
\begin{tabular}{ccc c}
\toprule
$bM$ & $\mueff^{2}M^{2}$ & $\Re(\Mw)$ & $\Im(\Mw)$ \\
\midrule
$-10^{-2}$        & $+2.0\times 10^{-3}$ & $0.27363389$ & $-0.09658956$ \\
$-5\times10^{-3}$ & $+1.0\times 10^{-3}$ & $0.27313411$ & $-0.09690897$ \\
$-2\times10^{-3}$ & $+4.0\times 10^{-4}$ & $0.27283436$ & $-0.09710040$ \\
$-10^{-3}$        & $+2.0\times 10^{-4}$ & $0.27273446$ & $-0.09716418$ \\
$0$               & $0$                  & $0.27263457$ & $-0.09722795$ \\
$+10^{-3}$        & $-2.0\times 10^{-4}$ & $0.27253470$ & $-0.09729169$ \\
$+2\times10^{-3}$ & $-4.0\times 10^{-4}$ & $0.27243483$ & $-0.09735542$ \\
$+5\times10^{-3}$ & $-1.0\times 10^{-3}$ & $0.27213529$ & $-0.09754650$ \\
$+10^{-2}$        & $-2.0\times 10^{-3}$ & $0.27163626$ & $-0.09786463$ \\
\bottomrule
\end{tabular}
\caption{Continuation of the fundamental counter-rotating mode
($\ell=1$, $m=-1$, $\mu M=0$) at $a=0.3M$, $qM=0.1$, as $bM$ is swept
through zero. The root varies smoothly and approximately linearly across the sign
crossing of $\mueff^{2}$ at $qbm=0$, and the two independently
implemented branch-selection prescriptions described in the text
return identical values at every displayed digit. No square-root
branch point of $\varrho$ is approached anywhere in the sweep.}
\label{tab.branch}
\end{table}

The outgoing condition \eqref{QNM.inf} is imposed at the formal
infinity of the extended equation, where it defines the auxiliary
Kerr spectrum. The inner validity scale \eqref{intermediate.region}
does not turn that condition into a finite-radius boundary condition
for the magnetized spacetime. Section~\ref{subsec.precision}
separates numerical precision from the local power counting and
the unresolved global comparison.

\section{Continued-fraction method and benchmark validation}
\label{sec.method}

\subsection{Numerical implementation}

In this subsection the recurrence is written in units $M=1$:
$r$, $a$, $\omega$ and mass parameters denote their dimensionless
counterparts $r/M$, $a/M$, $M\omega$ and $M\mu$. In dimensional
notation the power-law exponent below is
$\chi=M(\mueff^2-2\omega^2)/\varrho$.
The QNM spectrum is computed using the Leaver continued-fraction method
\cite{Leaver:1985ax} with Dolan's massive-scalar Kerr recurrence
\cite{Dolan:2007mj}. With Dolan's variable
\be
u=\frac{r-r_{+}}{r-r_{-}}
\label{Dolan.u}
\ee
and the radial-function ansatz
\be
R_{\ell m}=(r-r_{+})^{-i\sigma}(r-r_{-})^{i\sigma+\chi-1}
e^{\varrho r}\sum_{n=0}^{\infty}a_{n}u^{n},
\label{series.ansatz}
\ee
the Kerr massive-scalar radial equation gives the three-term recurrence
\be
\alpha_{n}a_{n+1}+\beta_{n}a_{n}+\gamma_{n}a_{n-1}=0,\qquad n\ge 1,
\label{recurrence}
\ee
with the explicit coefficients $\alpha_{n},\beta_{n},\gamma_{n}$ given by
Dolan \cite{Dolan:2007mj}. We denote Dolan's large-radius exponent by
$\varrho$ rather than $q$ to avoid confusion with the scalar charge. The
QNM condition is the standard minimal-solution condition,
\be
\beta_{0}-\frac{\alpha_{0}\gamma_{1}}
{\beta_{1}-\dfrac{\alpha_{1}\gamma_{2}}{\beta_{2}-\cdots}}=0.
\label{CF.eq}
\ee

The structural simplification of the reference gauge,
established in Sec.~\ref{sec.KG}, is that within the adopted
$\mathcal{O}(bq)$ truncation the weakly magnetized charged-scalar
radial problem can be cast in the Kerr massive-scalar form under the
mass substitution \eqref{mueff.def}. Dolan's recurrence
\eqref{recurrence}--\eqref{CF.eq} therefore applies as a
parameter-deformed scheme under this substitution, applied through the
Dolan parameters
\be
\varrho=\sqrt{\mueff^{2}-\omega^{2}},\qquad
\chi=\frac{\mueff^{2}-2\omega^{2}}{\varrho},\qquad
c^{2}=a^{2}(\omega^{2}-\mueff^{2}),
\label{Dolan.params.PLB}
\ee
the last entering through the angular eigenvalue \eqref{lambda.approx}.
For the positive-frequency damped branch considered here, the sign of
$\varrho$ is fixed by continuity from the massless Kerr limit,
$\varrho\to i\omega$ as $\mueff^{2}\to 0$. Throughout the parameter
range studied this is equivalent to selecting $\Re\varrho>0$ (the
half-plane selection rule referred to in Sec.~\ref{sec.QNM}); in the
continuation-tracked implementation, the sign at each parameter step
is instead chosen to minimize the distance from the preceding value of
$\varrho$.

To make the correspondence with Dolan's conventions fully explicit,
Table~\ref{tab.dictionary} lists the complete parameter dictionary.
The recurrence coefficients $\alpha_{n},\beta_{n},\gamma_{n}$ of
Ref.~\cite{Dolan:2007mj} are polynomials in $n$ built from five
constants $c_{0},\ldots,c_{4}$, which are explicit functions of
$\omega$, $\varrho$, $a$, $m$, and $\Lambda_{\ell m}$ (in the units
$M=1$). Inspection of those expressions shows that the scalar mass
enters them \emph{only} through $\varrho$ (equivalently $\chi$, since
$\chi=\varrho-\omega^{2}/\varrho$) and through
$\Lambda_{\ell m}(c^{2})$. Consequently, at $\mathcal{O}(bq)$ the
magnetic field enters the recurrence exclusively through the
substitution $\mu^{2}\to\mueff^{2}$ in $\varrho$, $\chi$, and
$\Lambda_{\ell m}(c^{2})$; the horizon exponent $\sigma$ is unmodified
(Sec.~\ref{sec.QNM}), the separation constant retains its unmagnetized
form as a function of $c^{2}$, and no coefficient acquires an
independent, term-by-term magnetic correction. The substitution
\eqref{mueff.def} is thus the \emph{only} $\mathcal{O}(bq)$
modification of Dolan's Kerr recurrence in the reference gauge at the order retained. The recurrence uses Dolan's coefficients verbatim (in the units
$M=1$ of Ref.~\cite{Dolan:2007mj}) with the parameter mapping of
Table~\ref{tab.dictionary}.

\begin{table}[htbp]
\centering
\footnotesize
\setlength{\tabcolsep}{4pt}
\renewcommand{\arraystretch}{1.4}
\begin{tabular}{lll}
\toprule
Quantity & Dolan \cite{Dolan:2007mj} & Present work \\
\midrule
mass-squared parameter & $\mu^{2}$ & $\mueff^{2}=\mu^{2}+2qbm$ \\
horizon exponent & $\sigma=\dfrac{2r_{+}(\omega-\omega_{c})}{r_{+}-r_{-}}$
 \;($M=1$) & identical; $\mueff^{2}$ does not enter \\
large-radius exponent & $q=\sqrt{\mu^{2}-\omega^{2}}$
 & $\varrho=\sqrt{\mueff^{2}-\omega^{2}}$ \\
auxiliary exponent & $\chi=\dfrac{\mu^{2}-2\omega^{2}}{q}$
 & $\chi=\dfrac{\mueff^{2}-2\omega^{2}}{\varrho}$ \\
spheroidicity & $c^{2}=a^{2}(\omega^{2}-\mu^{2})$
 & $c^{2}=a^{2}(\omega^{2}-\mueff^{2})$ \\
angular eigenvalue & $\Lambda_{\ell m}(c^{2})$
 & same expansion \eqref{lambda.approx}, with $c^{2}$ above \\
radial separation constant & $\Lambda_{\ell m}$, entering directly
 & $K_{\ell m}=\Lambda_{\ell m}(c^{2})$; no additional \\
 & in the radial equation & magnetic constant appears \\
\bottomrule
\end{tabular}
\caption{Dictionary between Dolan's notation \cite{Dolan:2007mj} and
the present work. Dolan's large-radius exponent is denoted $\varrho$
here to avoid confusion with the scalar charge $q$; the branch of
$\varrho$ is fixed by analytic continuation from the unmagnetized Kerr
QNM branch. The recurrence coefficients
$\alpha_{n},\beta_{n},\gamma_{n}$ depend on the scalar mass only
through $\varrho$, $\chi$, and $\Lambda_{\ell m}(c^{2})$, so the
mass substitution \eqref{mueff.def} is transmitted to the
continued fraction entirely through this table.}
\label{tab.dictionary}
\end{table}

The roots of \eqref{CF.eq} are obtained by Newton iteration in the complex
$\omega$-plane. For each parameter block $(a,\mu,m)$, the corresponding
$b=0$ Kerr root is used as the initial seed, and the root is then tracked
continuously through $bM=10^{-3}$ and $bM=10^{-2}$. At $b=0$ the scalar
charge drops out of the weak-field equation, since all charge-dependent
terms are proportional to $bq$. All production values in this paper use
truncation order $N=120$ ($N=300$ for the illustrative near-extremal
rows $a/M\ge 0.8$ of Tables~\ref{tab.Dolan2} and \ref{tab.Dolan3}), at
which the residual $N$-dependence is far below the displayed digits
throughout the surveyed range. This depth is dictated by the
slowest-converging corner of the survey, not by a favorable entry:
Table~\ref{tab.convergence} shows the $N$-ladder both at a
representative production point and at the most difficult surveyed
point, the massive counter-rotating mode at $a=0.5M$, $\mu M=0.3$,
$m=-1$. At the latter, the root still drifts by $\approx 7\times
10^{-6}$ between $N=50$ and $N=80$---above the rounding level of the
six-decimal benchmark tables---and settles only near $N\simeq 120$,
with a residual $N=120\to 200$ drift below $5\times 10^{-8}$. A
truncation depth adequate at co-rotating massless points is therefore
not uniformly adequate, and the production depth is set by the worst
case.

\begin{table}[htbp]
\centering
\begin{tabular}{c|cc|cc}
\toprule
& \multicolumn{2}{c|}{$(a,\mu M,m)=(0.3,0.3,+1)$}
& \multicolumn{2}{c}{$(a,\mu M,m)=(0.5,0.3,-1)$} \\
$N$ & $\Re(\Mw)$ & $\Im(\Mw)$ & $\Re(\Mw)$ & $\Im(\Mw)$ \\
\midrule
$50$  & 0.35599856 & -0.07564888 & 0.30929757 & -0.06369366 \\
$80$  & 0.35599867 & -0.07564897 & 0.30929142 & -0.06369104 \\
$120$ & 0.35599867 & -0.07564897 & 0.30929125 & -0.06369056 \\
$200$ & 0.35599867 & -0.07564897 & 0.30929125 & -0.06369052 \\
\bottomrule
\end{tabular}
\caption{Convergence of the magnetized continued-fraction root with the
truncation order $N$ at two points of the survey ($\ell=1$, $n=0$,
$qM=0.1$, $bM=10^{-2}$): a representative co-rotating production point
(left) and the slowest-converging surveyed point, the massive
counter-rotating mode at $a=0.5M$ (right). The co-rotating root is
converged to $\sim 10^{-7}$ already at $N=50$, whereas the difficult
root drifts by $\approx 7\times 10^{-6}$ between $N=50$ and $N=80$ and
settles only near $N=120$; the $N=120\to 200$ drift there is below
$5\times 10^{-8}$. The production depth $N=120$ is set by this worst
case.}
\label{tab.convergence}
\end{table}

\subsection{Validation against Dolan's benchmark}

The principal validation is obtained by setting $b=0$ and $q=0$. In this
limit the implementation reduces to the Kerr massive-scalar
continued-fraction problem, and Dolan's tabulated spectra
\cite{Dolan:2007mj} provide a stringent benchmark.
Tables~\ref{tab.Dolan2} and \ref{tab.Dolan3} compare the present code with
Dolan's Tables II and III for the counter-rotating mode $\ell=1$, $m=-1$
and the co-rotating mode $\ell=2$, $m=2$ respectively, over $\mu M\in
\{0,0.1,0.2,0.3\}$ and $a/M\in\{0,0.1,\ldots,0.9,0.95,0.99\}$. The
agreement with Dolan's six-decimal entries is close to the reference rounding level for the $\ell=1$ branch throughout $a\le 0.7M$: all $\ell=1$, $m=-1$
differences lie below $6\times 10^{-7}$, while the $\ell=2$, $m=+2$
differences remain below $3.1\times 10^{-6}$ at $a=0.7M$, the residual containing numerical and angular-approximation effects
that have not been isolated separately at these larger spins.
The near-extremal rows ($a/M\ge0.8$, computed at depth $N=300$)
are retained as illustrative comparisons. Their discrepancies are
not covered by the accuracy estimate for the magnetized survey,
which is restricted to $a/M\le0.5$.

\begin{table}[htbp]
\centering
\small
\begin{tabular}{cc|cc|cc}
\toprule
& & \multicolumn{2}{c|}{Dolan II ($\ell=1,m=-1$)} & \multicolumn{2}{c}{Dolan III ($\ell=2,m=+2$)} \\
$a/M$ & $\mu M$ & $|\Delta\Re(\Mw)|$ & $|\Delta\Im(\Mw)|$
                & $|\Delta\Re(\Mw)|$ & $|\Delta\Im(\Mw)|$ \\
\midrule
$0.0$ & $0$   & 1.33e-07  & 1.11e-08
              & 1.28e-07 & 2.24e-07 \\
$0.0$ & $0.3$ & 1.93e-07  & 2.90e-07
              & 3.74e-07 & 1.41e-07 \\
$0.3$ & $0$   & 4.26e-07  & 5.46e-08
              & 2.12e-07 & 4.29e-07 \\
$0.3$ & $0.3$ & 5.31e-09  & 2.87e-07
              & 3.55e-07 & 1.91e-08 \\
$0.5$ & $0$   & 2.31e-07  & 5.68e-07
              & 8.82e-08 & 7.57e-07 \\
$0.5$ & $0.3$ & 4.65e-07  & 3.69e-07
              & 9.16e-08 & 1.57e-07 \\
$0.7$ & $0$   & 3.61e-07  & 8.39e-08
              & 3.02e-06 & 2.04e-06 \\
$0.7$ & $0.3$ & 3.12e-07  & 5.49e-07
              & 2.38e-06 & 1.33e-06 \\
\bottomrule
\end{tabular}
\caption{Validation error budget against Dolan's tabulated values.
Absolute differences in $\Re(\Mw)$ and $\Im(\Mw)$ between the present
code (at $b=0$, $q=0$) and Dolan's Tables II and III, computed at
continued-fraction depth $N=120$ in 20--25-digit arithmetic against
Dolan's six-decimal tabulated values. All $\ell=1$, $m=-1$ entries
are below $6\times10^{-7}$, close to the rounding level of the
reference data across the full range $a\le 0.7M$; the residual few
$\times 10^{-6}$ in the $\ell=2$ column at $a=0.7M$ is not decomposed into radial and angular errors here. The magnetized
survey of Sec.~\ref{sec.results} stays at $a\le 0.5M$.}
\label{tab.errorbudget}
\end{table}

\begin{table}[htbp]
\centering
\small
\setlength{\tabcolsep}{3.5pt}
\begin{tabular}{c|cc|cc|cc|cc}
\toprule
& \multicolumn{2}{c|}{$\mu M=0$}
& \multicolumn{2}{c|}{$\mu M=0.1$}
& \multicolumn{2}{c|}{$\mu M=0.2$}
& \multicolumn{2}{c}{$\mu M=0.3$} \\
$a/M$ & $\Re$ & $\Im$ & $\Re$ & $\Im$ & $\Re$ & $\Im$ & $\Re$ & $\Im$ \\
\midrule
$0.0$  & $0.292936$ & $-0.097660$ & $0.297416$ & $-0.094957$ & $0.310957$ & $-0.086593$ & $0.333777$ & $-0.071658$ \\
$0.1$  & $0.285570$ & $-0.097626$ & $0.290234$ & $-0.094747$ & $0.304341$ & $-0.085845$ & $0.328135$ & $-0.069968$ \\
$0.2$  & $0.278833$ & $-0.097475$ & $0.283672$ & $-0.094427$ & $0.298318$ & $-0.085009$ & $0.323048$ & $-0.068231$ \\
$0.3$  & $0.272635$ & $-0.097228$ & $0.277641$ & $-0.094019$ & $0.292803$ & $-0.084108$ & $0.318436$ & $-0.066465$ \\
$0.4$  & $0.266901$ & $-0.096901$ & $0.272068$ & $-0.093537$ & $0.287726$ & $-0.083154$ & $0.314235$ & $-0.064680$ \\
$0.5$  & $0.261572$ & $-0.096506$ & $0.266893$ & $-0.092994$ & $0.283032$ & $-0.082158$ & $0.310392$ & $-0.062887$ \\
$0.6$  & $0.256596$ & $-0.096051$ & $0.262066$ & $-0.092399$ & $0.278671$ & $-0.081130$ & $0.306865$ & $-0.061091$ \\
$0.7$  & $0.251928$ & $-0.095547$ & $0.257544$ & $-0.091760$ & $0.274604$ & $-0.080076$ & $0.303615$ & $-0.059297$ \\
$0.8$  & $0.247531$ & $-0.095001$ & $0.253289$ & $-0.091085$ & $0.270795$ & $-0.079004$ & $0.300613$ & $-0.057513$ \\
$0.9$  & $0.243371$ & $-0.094423$ & $0.249270$ & $-0.090384$ & $0.267217$ & $-0.077920$ & $0.297831$ & $-0.055739$ \\
$0.95$  & $0.241372$ & $-0.094125$ & $0.247341$ & $-0.090026$ & $0.265506$ & $-0.077375$ & $0.296515$ & $-0.054858$ \\
$0.99$  & $0.239810$ & $-0.093884$ & $0.245834$ & $-0.089737$ & $0.264173$ & $-0.076940$ & $0.295495$ & $-0.054199$ \\
\midrule
\multicolumn{9}{l}{\small Dolan \cite{Dolan:2007mj} Table II (selected rows):} \\
$0.0$  & $0.292936$ & $-0.097660$ & $0.297416$ & $-0.094957$ & $0.310957$ & $-0.086593$ & $0.333777$ & $-0.071658$ \\
$0.3$  & $0.272635$ & $-0.097228$ & $0.277641$ & $-0.094019$ & $0.292803$ & $-0.084108$ & $0.318436$ & $-0.066465$ \\
$0.5$  & $0.261572$ & $-0.096505$ & $0.266893$ & $-0.092994$ & $0.283032$ & $-0.082158$ & $0.310392$ & $-0.062887$ \\
$0.7$  & $0.251928$ & $-0.095547$ & $0.257544$ & $-0.091760$ & $0.274604$ & $-0.080076$ & $0.303615$ & $-0.059298$ \\
$0.99$ & $0.239810$ & $-0.093882$ & $0.245834$ & $-0.089736$ & $0.264173$ & $-0.076939$ & $0.295520$ & $-0.054158$ \\
\bottomrule
\end{tabular}
\caption{QNM frequencies $\Mw$ for $\ell=1$, $m=-1$, $b=0$, $q=0$ (upper
block: present code; lower block: Dolan \cite{Dolan:2007mj} Table II).
Frequencies follow the time convention $\Phi\sim e^{-i\omega t}$, so
$\Im(\Mw)<0$ corresponds to a damped mode.
Agreement is within a few $\times10^{-6}$ for $a\le0.7M$. The illustrative $a/M\ge0.8$ rows use $N=300$ and show absolute discrepancies up to a few $\times10^{-5}$. Their separate radial and angular errors are unestablished, and they are not used in the magnetized survey.}
\label{tab.Dolan2}
\end{table}

\begin{table}[htbp]
\centering
\small
\setlength{\tabcolsep}{3.5pt}
\begin{tabular}{c|cc|cc|cc|cc}
\toprule
& \multicolumn{2}{c|}{$\mu M=0$}
& \multicolumn{2}{c|}{$\mu M=0.1$}
& \multicolumn{2}{c|}{$\mu M=0.2$}
& \multicolumn{2}{c}{$\mu M=0.3$} \\
$a/M$ & $\Re$ & $\Im$ & $\Re$ & $\Im$ & $\Re$ & $\Im$ & $\Re$ & $\Im$ \\
\midrule
$0.0$  & $0.483644$ & $-0.096759$ & $0.486804$ & $-0.095675$ & $0.496327$ & $-0.092389$ & $0.512346$ & $-0.086795$ \\
$0.1$  & $0.499482$ & $-0.096666$ & $0.502456$ & $-0.095674$ & $0.511419$ & $-0.092663$ & $0.526497$ & $-0.087528$ \\
$0.2$  & $0.517121$ & $-0.096382$ & $0.519901$ & $-0.095483$ & $0.528281$ & $-0.092755$ & $0.542378$ & $-0.088092$ \\
$0.3$  & $0.536979$ & $-0.095839$ & $0.539557$ & $-0.095036$ & $0.547326$ & $-0.092595$ & $0.560397$ & $-0.088418$ \\
$0.4$  & $0.559647$ & $-0.094931$ & $0.562011$ & $-0.094226$ & $0.569137$ & $-0.092083$ & $0.581127$ & $-0.088406$ \\
$0.5$  & $0.585990$ & $-0.093495$ & $0.588127$ & $-0.092891$ & $0.594568$ & $-0.091052$ & $0.605408$ & $-0.087895$ \\
$0.6$  & $0.617365$ & $-0.091246$ & $0.619257$ & $-0.090747$ & $0.624960$ & $-0.089225$ & $0.634560$ & $-0.086608$ \\
$0.7$  & $0.656102$ & $-0.087651$ & $0.657724$ & $-0.087261$ & $0.662616$ & $-0.086070$ & $0.670850$ & $-0.084019$ \\
$0.8$  & $0.706830$ & $-0.081524$ & $0.708144$ & $-0.081249$ & $0.712108$ & $-0.080410$ & $0.718781$ & $-0.078964$ \\
$0.9$  & $0.781653$ & $-0.069295$ & $0.782584$ & $-0.069147$ & $0.785393$ & $-0.068696$ & $0.790124$ & $-0.067916$ \\
$0.95$  & $0.841005$ & $-0.056478$ & $0.841675$ & $-0.056402$ & $0.843697$ & $-0.056170$ & $0.847105$ & $-0.055769$ \\
$0.99$  & $0.928055$ & $-0.031069$ & $0.928380$ & $-0.031060$ & $0.929361$ & $-0.031031$ & $0.931016$ & $-0.030982$ \\
\midrule
\multicolumn{9}{l}{\small Dolan \cite{Dolan:2007mj} Table III (selected rows):} \\
$0.0$  & $0.483644$ & $-0.096759$ & $0.486804$ & $-0.095675$ & $0.496327$ & $-0.092389$ & $0.512346$ & $-0.086795$ \\
$0.3$  & $0.536979$ & $-0.095839$ & $0.539557$ & $-0.095036$ & $0.547326$ & $-0.092595$ & $0.560397$ & $-0.088418$ \\
$0.5$  & $0.585990$ & $-0.093494$ & $0.588127$ & $-0.092890$ & $0.594568$ & $-0.091052$ & $0.605408$ & $-0.087895$ \\
$0.7$  & $0.656099$ & $-0.087649$ & $0.657722$ & $-0.087259$ & $0.662614$ & $-0.086068$ & $0.670848$ & $-0.084018$ \\
$0.99$ & $0.928028$ & $-0.031063$ & $0.928353$ & $-0.031054$ & $0.929336$ & $-0.031026$ & $0.930994$ & $-0.030977$ \\
\bottomrule
\end{tabular}
\caption{QNM frequencies $\Mw$ for $\ell=2$, $m=2$, $b=0$, $q=0$ (upper
block: present code; lower block: Dolan \cite{Dolan:2007mj} Table III).
Frequencies follow the time convention $\Phi\sim e^{-i\omega t}$, so
$\Im(\Mw)<0$ corresponds to a damped mode.
Agreement is within a few $\times10^{-6}$ for $a\le0.7M$. The illustrative $a/M\ge0.8$ rows use $N=300$ and show absolute discrepancies up to a few $\times10^{-5}$. Their separate radial and angular errors are unestablished, and they are not used in the magnetized survey.}
\label{tab.Dolan3}
\end{table}

For the co-rotating $\ell=1$, $m=+1$ mode used in the magnetized survey
below, Dolan's Tables II and III do not provide the corresponding entries. The corresponding $b=0$, $q=0$
spectra obtained with the same validated implementation are listed in
Table~\ref{tab.Cp}. These are the $b=0$ reference values for the
magnetized continuation in Table~\ref{tab.QNM.MK}.

\begin{table}[htbp]
\centering
\small
\setlength{\tabcolsep}{3.5pt}
\begin{tabular}{c|cc|cc|cc|cc}
\toprule
& \multicolumn{2}{c|}{$\mu M=0$}
& \multicolumn{2}{c|}{$\mu M=0.1$}
& \multicolumn{2}{c|}{$\mu M=0.2$}
& \multicolumn{2}{c}{$\mu M=0.3$} \\
$a/M$ & $\Re$ & $\Im$ & $\Re$ & $\Im$ & $\Re$ & $\Im$ & $\Re$ & $\Im$ \\
\midrule
$0.0$  & $0.292936$ & $-0.097660$ & $0.297416$ & $-0.094957$ & $0.310957$ & $-0.086593$ & $0.333777$ & $-0.071658$ \\
$0.1$  & $0.301045$ & $-0.097547$ & $0.305329$ & $-0.095029$ & $0.318274$ & $-0.087228$ & $0.340076$ & $-0.073277$ \\
$0.2$  & $0.310043$ & $-0.097245$ & $0.314119$ & $-0.094920$ & $0.326433$ & $-0.087709$ & $0.347162$ & $-0.074790$ \\
$0.3$  & $0.320126$ & $-0.096691$ & $0.323981$ & $-0.094569$ & $0.335621$ & $-0.087979$ & $0.355212$ & $-0.076147$ \\
$0.4$  & $0.331567$ & $-0.095792$ & $0.335181$ & $-0.093883$ & $0.346095$ & $-0.087950$ & $0.364470$ & $-0.077267$ \\
$0.5$  & $0.344753$ & $-0.094395$ & $0.348105$ & $-0.092714$ & $0.358230$ & $-0.087478$ & $0.375284$ & $-0.078022$ \\
$0.6$  & $0.360285$ & $-0.092243$ & $0.363346$ & $-0.090806$ & $0.372595$ & $-0.086321$ & $0.388193$ & $-0.078188$ \\
$0.7$  & $0.379159$ & $-0.088849$ & $0.381888$ & $-0.087679$ & $0.390142$ & $-0.084014$ & $0.404090$ & $-0.077335$ \\
\bottomrule
\end{tabular}
\caption{Reference $b=0$, $q=0$ QNM frequencies $\Mw$ for the co-rotating
mode $\ell=1$, $m=+1$, computed with the same validated continued-fraction
implementation. Frequencies follow the time convention
$\Phi\sim e^{-i\omega t}$, so $\Im(\Mw)<0$ corresponds to a damped
mode. These values seed the magnetized continuation in
Table~\ref{tab.QNM.MK}.}
\label{tab.Cp}
\end{table}

\subsection{Numerical precision of the magnetized frequencies}
\label{subsec.precision}

Three distinct questions must be kept separate: numerical accuracy
within the auxiliary problem; the size of omitted terms in the
local differential operator; and whether any chosen global
magnetized resonance is approximated by an auxiliary root. Only
the first is tested numerically in the spectrum tables.

\emph{Numerical error} $\delta\omega_{\rm num}$. All roots are
computed in 20--25-digit arithmetic, so floating-point error is
negligible. Three sources remain. (i)
\emph{Truncation of the continued fraction}: at the production depth
$N=120$ the root drift relative to $N=200$ is below $5\times 10^{-8}$
even at the slowest-converging surveyed point
(Table~\ref{tab.convergence}); the quoted uncertainty is the largest
observed $N$-drift over the survey, not that of a favorable entry.
(ii) \emph{Approximate angular eigenvalue}: replacing the leading
Seidel approximation \eqref{lambda.approx} by the next-order
expansion, or by a direct numerical solution of the spheroidal
eigenvalue problem, changes the roots by less than $3\times 10^{-7}$
at all sampled points (Tables~\ref{tab.lambda.HO} and
\ref{tab.lambda.spec}, Sec.~\ref{subsec.lambda.HO}). (iii)
\emph{Accuracy of the unmagnetized backbone}: agreement with Dolan's
six-decimal reference values is below $6\times 10^{-7}$ for the
$\ell=1$ backbone across $a\le 0.7M$
(Table~\ref{tab.errorbudget}), consistent with the rounding floor of
the reference data. Combining (i)--(iii), we estimate
$\delta\omega_{\rm num}\lesssim 1\times 10^{-6}$ in $\Mw$ over the
magnetized survey; all four decimals displayed in
Tables~\ref{tab.QNM.MK} and \ref{tab.l2} are significant \emph{as
solutions of the truncated spectral problem}.

\emph{Local operator truncation.} For a neutral Kerr seed,
$g_{\mu\nu}=g^{(0)}_{\mu\nu}+b^2g^{(2)}_{\mu\nu}+\cdots$ and
$A_\mu=b\mathcal A^{(1)}_\mu+b^3\mathcal A^{(3)}_\mu+\cdots$.
There is consequently no $b^2q$ term. Define dimensionless
$\beta=bM$ and $e=qM$. At fixed $r/M$ of order unity, the retained
interaction scales as $e\beta$, and the first omitted operator
contributions carry factors
\be
\beta^2,\qquad e^2\beta^2,\qquad e\beta^3.
\label{operator.powercounting}
\ee
For $(e,\beta)=(0.1,10^{-2})$ these factors are
$10^{-4}$, $10^{-6}$ and $10^{-7}$, respectively, compared with
$e\beta=10^{-3}$. Their coefficients depend on position and mode
parameters. These are dimensionless \emph{operator power-counting
factors}, not computed corrections to $M\omega$. The ratio
$|\beta/e|$ compares formal local orders; it is not a relative
frequency error. In particular, it supplies neither a ten-percent
nor a one-percent accuracy statement for a magnetized resonance.
The hierarchy $|e|\gg|\beta|$ only favors the retained local
charge coupling over the first metric correction, subject to
coefficients and the radial restrictions in
Eq.~\eqref{intermediate.region}.

\emph{Global spectral comparison.} No exact outer boundary problem
or matching calculation has been supplied. Therefore an error
$\omega_{\rm exact}-\omega_{\rm aux}$ is not evaluated, and no
percentage bound is assigned to it. The sizes in
Eq.~\eqref{operator.powercounting} cannot be promoted to such a
bound: a change in distant boundary conditions can affect a
resonance even if the inner equation changes little. This limitation
is separate from convergence of the continued fraction.

The observed magnetic shifts must be compared against the
\emph{numerical} error floor $\delta\omega_{\rm num}$.
The smallest shifts in the survey occur at $bM=10^{-3}$, where
$|\delta_{\rm mag}\Re(\Mw)|\simeq (0.7$--$1.1)\times 10^{-4}$ (approximately one
tenth of the $bM=10^{-2}$ shifts of Table~\ref{tab.slope}, by
linearity); this exceeds the estimated error floor by more than an
order of magnitude. At $bM=10^{-2}$ the shifts,
$|\delta_{\rm mag}\Re(\Mw)|\simeq (0.7$--$1.1)\times 10^{-3}$, exceed
the floor by over two orders of magnitude. Every magnetic shift
displayed in this paper is therefore numerically significant, and in
particular larger than the angular-eigenvalue error at every point of
the survey. The individual shifts at $bM=10^{-2}$ are resolved to six
decimals in the column $\delta_{\rm mag}^{\rm obs}$ of
Table~\ref{tab.slope}; by the linearity verified in
Table~\ref{tab.linearity}, the $bM=10^{-3}$ shifts are approximately one tenth of
those values, and are therefore individually resolved internally even
where the four-decimal display of Table~\ref{tab.QNM.MK} rounds them
to a single unit in the last digit. Here ``numerically significant'' means resolved within the auxiliary
problem. It does not quantify how well any of these roots represents
a global magnetized mode.

\section{Magnetic corrections to the effective Kerr QNM spectrum}
\label{sec.results}

We now restrict to the fundamental $\ell=1$, $n=0$ modes at $a/M\in
\{0.3,0.5\}$, fixed scalar charge $qM=0.1$, scalar masses
$\mu M\in\{0,0.3\}$, and magnetic-field strengths $bM\in\{0,10^{-3},10^{-2}\}$,
for both co-rotating ($m=+1$) and counter-rotating ($m=-1$) sectors.

\subsection{Spectral results}

Table~\ref{tab.QNM.MK} presents the fundamental auxiliary effective Kerr spectrum.
The trends with $b$ are systematic and the magnetic shifts are
\emph{opposite in sign} between the two rotating sectors of equal
$|m|$: increasing $b$ shifts $\Re(\Mw)$ upward for $m=+1$ and downward
for $m=-1$, with the shifts approximately linear in $b$ over the displayed range;
all displayed shifts exceed the numerical error floor established in
Sec.~\ref{subsec.precision} by more than an order of magnitude.
The two magnitudes are not in general equal because $\Re(\Mw)$ is a
nonlinear function of $\mu^{2}$ whose slope depends on $m$ through the
underlying Kerr spectrum; we return to this in
Sec.~\ref{subsec.slope}. The imaginary parts also shift, with
$|\Im(\Mw)|$ moving in opposite directions in the two sectors,
modifying the damping time $\tau/M=1/|\Im(\Mw)|$ and the quality
factor $Q=\Re(\Mw)/[2|\Im(\Mw)|]$ accordingly.

The opposite-sign pattern is a direct consequence of the
sign-dependence of the asymptotic effective-mass shift,
$\mueff^{2}-\mu^{2}=2qbm$, which raises the effective mass for $m=+1$
and lowers it for $m=-1$. Within the adopted $\mathcal{O}(bq)$
truncation, the mass substitution $\mu^{2}\to\mueff^{2}$ is the only
$b$-dependence in the spectral problem; the magnitude of each shift
is set by the unmagnetized slope $\partial\Re(\Mw)/\partial(\mu M)^{2}$
evaluated at the corresponding $(a,m,\mu)$, multiplied by $2qbm M^{2}$.
The slopes for $m=+1$ and $m=-1$ generally differ, producing the
sector-dependent magnitudes seen in Table~\ref{tab.QNM.MK}.

\begin{table}[htbp]
\centering
\begin{tabular}{llllcccc}
\toprule
$a/M$ & $\mu M$ & $m$ & $bM$ &
$\Re(\Mw)$ & $\Im(\Mw)$ & $\tau/M$ & $Q$ \\
\midrule
\multicolumn{8}{l}{\textit{$a/M=0.3$, $\ell=1$, $n=0$, $qM=0.1$}} \\
\midrule
$0.3$ & $0$   & $+1$ & $0$        & 0.3201 & -0.0967 & 10.342 & 1.655 \\
$0.3$ & $0$   & $+1$ & $10^{-3}$  & 0.3202 & -0.0966 & 10.347 & 1.657 \\
$0.3$ & $0$   & $+1$ & $10^{-2}$  & 0.3209 & -0.0963 & 10.387 & 1.667 \\
\cmidrule(lr){3-8}
$0.3$ & $0$   & $-1$ & $0$        & 0.2726 & -0.0972 & 10.285 & 1.402 \\
$0.3$ & $0$   & $-1$ & $10^{-3}$  & 0.2725 & -0.0973 & 10.278 & 1.401 \\
$0.3$ & $0$   & $-1$ & $10^{-2}$  & 0.2716 & -0.0979 & 10.218 & 1.388 \\
\midrule
$0.3$ & $0.3$ & $+1$ & $0$        & 0.3552 & -0.0761 & 13.133 & 2.332 \\
$0.3$ & $0.3$ & $+1$ & $10^{-3}$  & 0.3553 & -0.0761 & 13.141 & 2.334 \\
$0.3$ & $0.3$ & $+1$ & $10^{-2}$  & 0.3560 & -0.0756 & 13.219 & 2.353 \\
\cmidrule(lr){3-8}
$0.3$ & $0.3$ & $-1$ & $0$        & 0.3184 & -0.0665 & 15.046 & 2.396 \\
$0.3$ & $0.3$ & $-1$ & $10^{-3}$  & 0.3183 & -0.0665 & 15.029 & 2.392 \\
$0.3$ & $0.3$ & $-1$ & $10^{-2}$  & 0.3174 & -0.0672 & 14.881 & 2.362 \\
\midrule
\multicolumn{8}{l}{\textit{$a/M=0.5$, $\ell=1$, $n=0$, $qM=0.1$}} \\
\midrule
$0.5$ & $0$   & $+1$ & $0$        & 0.3448 & -0.0944 & 10.594 & 1.826 \\
$0.5$ & $0$   & $+1$ & $10^{-3}$  & 0.3448 & -0.0944 & 10.598 & 1.827 \\
$0.5$ & $0$   & $+1$ & $10^{-2}$  & 0.3454 & -0.0941 & 10.631 & 1.836 \\
\cmidrule(lr){3-8}
$0.5$ & $0$   & $-1$ & $0$        & 0.2616 & -0.0965 & 10.362 & 1.355 \\
$0.5$ & $0$   & $-1$ & $10^{-3}$  & 0.2615 & -0.0966 & 10.355 & 1.354 \\
$0.5$ & $0$   & $-1$ & $10^{-2}$  & 0.2605 & -0.0972 & 10.288 & 1.340 \\
\midrule
$0.5$ & $0.3$ & $+1$ & $0$        & 0.3753 & -0.0780 & 12.817 & 2.405 \\
$0.5$ & $0.3$ & $+1$ & $10^{-3}$  & 0.3754 & -0.0780 & 12.823 & 2.407 \\
$0.5$ & $0.3$ & $+1$ & $10^{-2}$  & 0.3760 & -0.0776 & 12.883 & 2.422 \\
\cmidrule(lr){3-8}
$0.5$ & $0.3$ & $-1$ & $0$        & 0.3104 & -0.0629 & 15.901 & 2.468 \\
$0.5$ & $0.3$ & $-1$ & $10^{-3}$  & 0.3103 & -0.0630 & 15.880 & 2.464 \\
$0.5$ & $0.3$ & $-1$ & $10^{-2}$  & 0.3093 & -0.0637 & 15.701 & 2.428 \\
\bottomrule
\end{tabular}
\caption{Fundamental auxiliary effective-Kerr QNM frequencies $\Mw$ for $\ell=1$, $n=0$, $qM=0.1$,
in the reference gauge. The $b=0$ rows match
Tables~\ref{tab.Dolan2}, \ref{tab.Cp}. The damping time and quality factor
are $\tau/M=1/|\Im(\Mw)|$ and $Q=\Re(\Mw)/[2|\Im(\Mw)|]$. The principal
qualitative feature is the opposite-sign pattern in $m$ at fixed
$(a,\mu)$: $\Re(\Mw)$ shifts up for $m=+1$ and down for $m=-1$, both
to leading order in $b$, with sector-dependent magnitudes set by the
unmagnetized slope $\partial\Re(\Mw)/\partial(\mu M)^{2}$
(Sec.~\ref{subsec.slope}). Frequencies follow the time convention
$\Phi\sim e^{-i\omega t}$, so $\Im(\Mw)<0$ corresponds to a damped
mode; $Q$ denotes the quality factor, not a charge. All displayed
magnetic shifts exceed the numerical error floor of
Sec.~\ref{subsec.precision} by more than an order of magnitude; the
shifts at $bM=10^{-2}$ are resolved to six decimals in
Table~\ref{tab.slope}.}
\label{tab.QNM.MK}
\end{table}

Because the retained equation depends on $q$ and $b$ only through the
product $qb$ (times $m$), the spectrum obeys two exact reflection
identities, which we verified numerically as consistency checks of the
implementation. First, the trivial invariance
$\omega_{\ell m}(q,b)=\omega_{\ell m}(-q,-b)$ holds identically (the
two runs return the same root to all computed digits). Second, the
standard Kerr conjugation symmetry combines with the $qbm$ structure
into the field-reversal identity
\be
\omega_{\ell,-m}(q,b)\;=\;-\bigl[\omega_{\ell,m}(q,-b)\bigr]^{*},
\label{mirror.identity}
\ee
which relates the two rotating sectors at opposite field orientations.
We verified \eqref{mirror.identity} explicitly at
$(a,\mu M,qM,|bM|)=(0.3,0.3,0.1,10^{-2})$: the point
$-[\omega_{\ell,+1}(q,-b)]^{*}$ is reproduced as a root of the $m=-1$
continued-fraction equation to better than $10^{-20}$. Note that the
mirror image lies on the $\Re\,\omega<0$ branch: the two
\emph{positive}-frequency fundamentals tabulated for $m=\pm 1$ at the
\emph{same} field $b$ are independent modes, not conjugate images of
one another, which is why their magnetic shifts have unequal
magnitudes. Under the triple reversal $(q,b,m)\to(-q,-b,-m)$, the product
$qb$ is unchanged but $qbm$ changes sign. This is not a same-frequency
invariance. The conjugation identity \eqref{mirror.identity} relates
opposite real-frequency branches after the required field reversal;
it does not identify the two positive-frequency rows at fixed $b$.

\subsection{Effective-potential diagnostics}

The diagnostics of this subsection are intended as a \emph{qualitative
visualization} of how the magnetic term deforms the radial problem;
they do not constitute an independent quantitative explanation of the
spectral shifts, whose sign and magnitude follow from the mass-slope
analysis of Sec.~\ref{subsec.slope}. Two well-known caveats limit any
potential-based inference for QNMs: the frequency-dependence of
$V_{\rm eff}$ (the potential is evaluated at a fixed real reference
frequency, while the mode frequency is complex), and the fact that
complex QNM frequencies are not determined by the barrier height alone.
With this understanding, Figures~\ref{fig.Veff.mu0} and
\ref{fig.Veff.mu03} display the effective
potential $V_{\rm eff}(r)$ from \eqref{Veff} and its magnetic correction
$\Delta V_{\rm eff}(r)\equiv V_{\rm eff}^{\mathcal{O}(bq)}(r)-V_{\rm eff}^{\rm Kerr}(r)$
for the co-rotating mode $\ell=m=1$ at $a=0.3M$, $qM=0.1$, $bM=10^{-2}$,
evaluated at the real parts of the corresponding unmagnetized co-rotating
Kerr frequencies in Table~\ref{tab.Cp} ($\Re(\Mw)\simeq 0.3201$ for
$\mu M=0$ and $\Re(\Mw)\simeq 0.3552$ for $\mu M=0.3$).

At a fixed real reference frequency the angular eigenvalue must
also be varied with the effective mass. With the leading angular
approximation used here, $K_{11}=2-c^2/5$, the difference is
\be
\delta K_{11}=\frac{a^2}{5}(2qbm),\qquad
\Delta V_{\rm eff}(r)=-\frac{\Delta}{(r^2+a^2)^2}
\left[2qbm\,r^2+\delta K_{11}\right].
\label{potential.shift.complete}
\ee
The figures and Table~\ref{tab.Veffdiag} include both terms.
They use $M\omega_{\rm ref}=0.3201265$ for $\mu M=0$ and
$0.355212$ for $\mu M=0.3$, held fixed while $b$ changes.
The correction vanishes at the horizon and tends to $-2qbm$
at the formal auxiliary infinity. It is monotone over the plotted
range for the parameters used. At this angular order the difference
curve is independent of the chosen bare mass and real reference
frequency, although the individual potentials are not.

Since two sign conventions are in play, we spell them out step by
step. Equation \eqref{Schrodinger.form} is written in the form
$d^{2}Y/dy^{2}+V_{\rm eff}\,Y=0$, so $V_{\rm eff}$ contains the
frequency: at large radius $V_{\rm eff}\to\omega^{2}-\mueff^{2}$.
Comparing with the standard Schr\"odinger form
$d^{2}Y/dy^{2}+[\omega^{2}-U(r)]\,Y=0$ identifies the WKB barrier
\be
U(r)\;\equiv\;\omega^{2}-V_{\rm eff}(r),
\label{U.def}
\ee
which approaches the formal large-$r$ asymptote $U\to\mueff^{2}$ of
the analytically extended truncated equation (the physical Melvin far
region is not described by the truncation) and has the familiar
peaked-barrier shape in the photon-sphere region. A
\emph{negative} magnetic correction to $V_{\rm eff}$,
$\Delta V_{\rm eff}<0$
(Figs.~\ref{fig.Veff.mu0}--\ref{fig.Veff.mu03}), therefore corresponds
to a \emph{positive} shift of the barrier,
$\Delta U=-\Delta V_{\rm eff}>0$: the barrier becomes taller for the
co-rotating sector, and shorter for the counter-rotating one, where
the sign of $\Delta V_{\rm eff}$ reverses. This deformation is nothing
but the pictorial counterpart of the effective-mass shift: the barrier
of the extended truncated equation approaches $U\to\mueff^{2}$, so
raising $\mueff^{2}$ lifts the large-$r$ plateau and the outer flank
of the barrier. The sign of the
frequency shift itself is \emph{not} inferred from the barrier
picture; it follows quantitatively from the positivity of the
unmagnetized mass slope $\partial\Re(\Mw)/\partial(\mu M)^{2}>0$
together with the sign of $\mueff^{2}-\mu^{2}=2qbm$
(Sec.~\ref{subsec.slope}). The figures merely illustrate that the
direction of the barrier deformation is consistent with that
mass-slope mechanism.

Table~\ref{tab.Veffdiag} quantifies the magnetic shift of $V_{\rm eff}$
at the location $r_{\min}$ of the (Kerr) extremum of $V_{\rm eff}$ in the
photon-sphere region, $r_{\min}/M\simeq 2.68$--$2.92$, and at the reference
radius $r=4M$. The tabulated potential differences are $M^2\Delta V_{\rm eff}$. Both quantities scale linearly with $bq$, as expected at
leading order in the weak-field expansion.

\begin{figure}[htbp]
\centering
\begin{subfigure}[b]{0.48\textwidth}
\includegraphics[width=\textwidth]{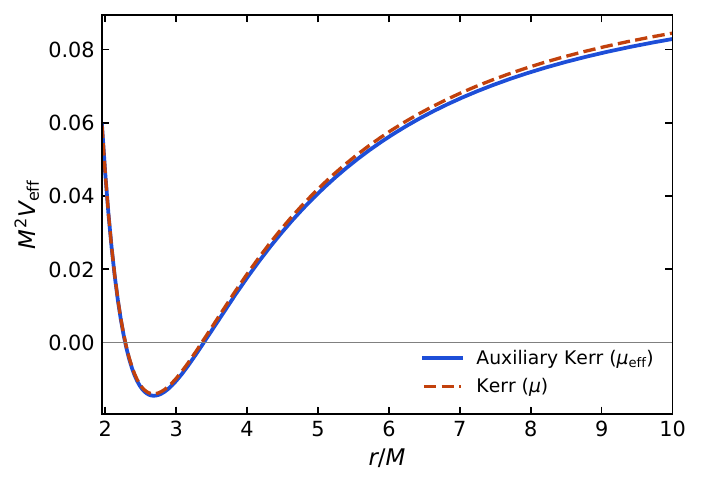}
\caption{$V_{\rm eff}(r)$, $\mu M=0$.}
\end{subfigure}
\hfill
\begin{subfigure}[b]{0.48\textwidth}
\includegraphics[width=\textwidth]{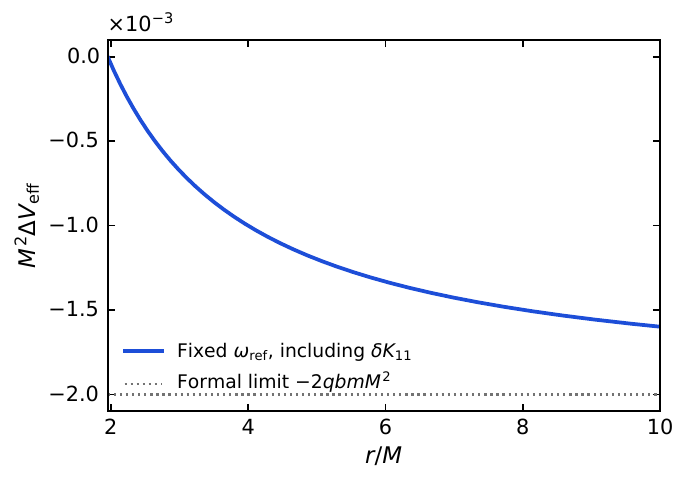}
\caption{$\Delta V_{\rm eff}(r)$, $\mu M=0$.}
\end{subfigure}
\caption{Effective potential and magnetic correction for the co-rotating
mode $\ell=m=1$ with $a=0.3M$, $\mu M=0$, $\Re(\Mw)\simeq 0.3201$,
$qM=0.1$, $bM=10^{-2}$, in the reference gauge. Left:
solid blue is the potential of the truncated $\mathcal{O}(bq)$
auxiliary equation, including the shifted angular eigenvalue of Eq.~\eqref{potential.shift.complete};
dashed red is unmagnetized Kerr. Right: the difference
$\Delta V_{\rm eff}^{\mathcal{O}(bq)}=V_{\rm eff}^{\mathcal{O}(bq)}-
V_{\rm eff}^{\rm Kerr}$, monotonic in $r$ and approaching the constant
$-2qbm$, the formal large-$r$ asymptote of the extended truncated
equation (dotted gray line).}
\label{fig.Veff.mu0}
\end{figure}

\begin{figure}[htbp]
\centering
\begin{subfigure}[b]{0.48\textwidth}
\includegraphics[width=\textwidth]{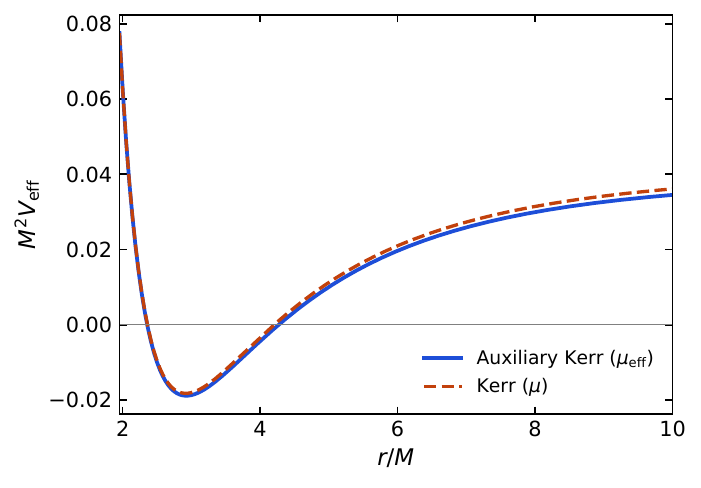}
\caption{$V_{\rm eff}(r)$, $\mu M=0.3$.}
\end{subfigure}
\hfill
\begin{subfigure}[b]{0.48\textwidth}
\includegraphics[width=\textwidth]{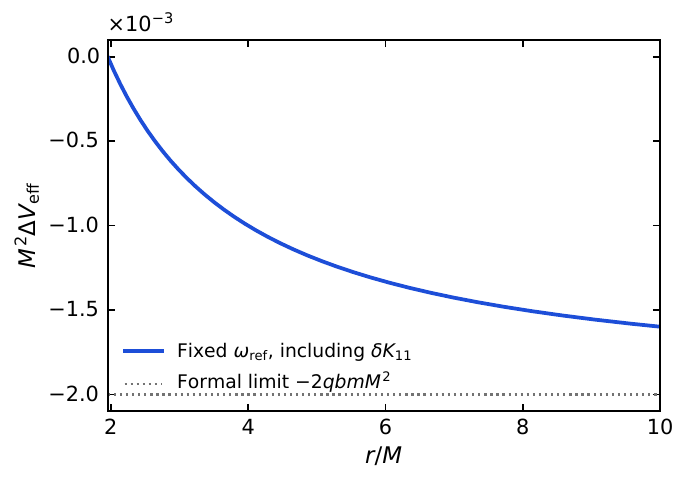}
\caption{$\Delta V_{\rm eff}(r)$, $\mu M=0.3$.}
\end{subfigure}
\caption{Same as Fig.~\ref{fig.Veff.mu0} but for $\mu M=0.3$,
$\Re(\Mw)\simeq 0.3552$. The barrier is taller and broader than at
$\mu M=0$; the qualitative shape and sign of $\Delta V_{\rm eff}$ are
unchanged.}
\label{fig.Veff.mu03}
\end{figure}

\begin{table}[htbp]
\centering
\begin{tabular}{llccc}
\toprule
$\mu M$ & $bM$ & $r_{\min}/M$ &
$M^2\Delta V_{\rm eff}(r_{\min})$ & $M^2\Delta V_{\rm eff}(4M)$ \\
\midrule
$0$ & $10^{-3}$ & $2.68052$ & $-5.210\times10^{-5}$ & $-1.001\times10^{-4}$ \\
$0$ & $5\times10^{-3}$ & $2.68052$ & $-2.605\times10^{-4}$ & $-5.005\times10^{-4}$ \\
$0$ & $10^{-2}$ & $2.68052$ & $-5.210\times10^{-4}$ & $-1.001\times10^{-3}$ \\
\midrule
$0.3$ & $10^{-3}$ & $2.91845$ & $-6.384\times10^{-5}$ & $-1.001\times10^{-4}$ \\
$0.3$ & $5\times10^{-3}$ & $2.91845$ & $-3.192\times10^{-4}$ & $-5.005\times10^{-4}$ \\
$0.3$ & $10^{-2}$ & $2.91845$ & $-6.384\times10^{-4}$ & $-1.001\times10^{-3}$ \\
\bottomrule
\end{tabular}
\caption{Effective-potential diagnostics for the co-rotating mode
$\ell=m=1$ at $a=0.3M$, $qM=0.1$. $r_{\min}$ is the location of the Kerr
extremum of $V_{\rm eff}$ in the photon-sphere region;
$\Delta V_{\rm eff}(r_{\min})$ is the value of the magnetic correction at
that point; $\Delta V_{\rm eff}(4M)$ is a reference value at $r=4M$. Both
$\Delta V_{\rm eff}(r_{\min})$ and $\Delta V_{\rm eff}(4M)$ are negative
in the convention of \eqref{Schrodinger.form} and scale linearly with
$bq$. A negative shift of $V_{\rm eff}$ corresponds to a positive shift
of the WKB barrier $U=\omega^{2}-V_{\rm eff}$; as discussed in the text,
this is the pictorial counterpart of the effective-mass shift and is
displayed as a qualitative illustration, not as an independent
derivation of the frequency shifts.}
\label{tab.Veffdiag}
\end{table}

\subsection{Consistency with the unmagnetized mass slope}
\label{subsec.slope}

Within the adopted $\mathcal{O}(bq)$ truncation, the
mass substitution $\mu^{2}\to\mueff^{2}=\mu^{2}+2qbm$ is the only
$b$-dependence in the auxiliary problem. In dimensionless variables
$s=(\mu M)^2$ and $\epsilon=2(qM)(bM)m$, let
$\Omega_{\ell mn}(a/M,s)$ denote a Kerr QNM branch. Then
\be
M\omega_{\rm aux}(b)=\Omega_{\ell mn}(a/M,s+\epsilon),\qquad
\left.\frac{\partial(M\omega_{\rm aux})}{\partial(bM)}\right|_{b=0}
=2(qM)m\,\partial_s\Omega_{\ell mn}(a/M,s).
\label{auxiliary.response}
\ee
The equality holds for the defined auxiliary spectrum on the same
continued branch (and for a numerical angular approximation applied
consistently on both sides). This is an \emph{analytic}
statement, established in Sec.~\ref{sec.KG}; consequently, the
complex magnetic shift at fixed $(a,\mu,m,\ell)$ must satisfy
\be
\delta_{\rm mag}(\Mw) \;=\; \left[\frac{\partial(\Mw)}{\partial(\mu M)^{2}}
\biggm|_{a,m,\ell}\right]\, 2qbm M^{2}
\;+\;\mathcal{O}(\epsilon^{2}),
\label{slope.prediction}
\ee
where $\epsilon=2(qM)(bM)m$, in both real and imaginary parts, with the complex slope evaluated on
the unmagnetized Kerr massive-scalar spectrum. Since
\eqref{slope.prediction} is guaranteed by the algebra of the truncated
equation, its numerical verification is an \emph{internal consistency
check of the implementation}---of the substitution's propagation
through the recurrence coefficients, the branch selection, and the
root finding---rather than an independent validation of the physical
picture; conversely, any statistically significant violation would
require checking the implementation, the finite-difference estimate and nonlinear terms. Solving the auxiliary equation at finite $b$ includes curvature in $\epsilon$; those terms are not the complete second-order magnetic correction to the original spacetime. The check remains informative in two
structural respects: the shifts have opposite signs for the sampled branches, whose real mass slopes have the same sign; their magnitude is set entirely by the unmagnetized
slope, which depends on $m$, so the magnitudes in the two rotating
sectors are not in general equal.

To carry out the check quantitatively, we extract the complex slope
from the unmagnetized spectra by finite differences on the
$(\mu M)^{2}$ grid supplied by Tables~\ref{tab.Dolan2} ($m=-1$) and
\ref{tab.Cp} ($m=+1$). At $\mu M=0$, we use the forward estimator
\be
\frac{\partial(\Mw)}{\partial(\mu M)^{2}}\biggm|_{a,m,\ell,\mu M=0}
\;\simeq\;
\frac{(\Mw)\big|_{\mu M=0.1}-(\Mw)\big|_{\mu M=0}}{0.01},
\label{slope.fd}
\ee
applied to real and imaginary parts alike. At $\mu M=0.3$, where the
curvature of $\Mw$ in $(\mu M)^{2}$ is no longer negligible at the
$10^{-5}$ level, we use the three-point one-sided estimator built on
the nodes $(\mu M)^{2}\in\{0.01,0.04,0.09\}$, i.e.\ the derivative at
$(\mu M)^{2}=0.09$ of the quadratic interpolant through the three
unmagnetized values. The predicted shift is
$\delta_{\rm mag}^{\rm pred}=
[\partial(\Mw)/\partial(\mu M)^{2}]\cdot 2qbm M^{2}$, and the
observed shift $\delta_{\rm mag}^{\rm obs}$ is the difference between
the magnetized and unmagnetized roots at the same point.
Table~\ref{tab.slope} reports the comparison for both parts of $\Mw$.
The agreement is at the level of a few $\times 10^{-6}$ in
$\Re(\Mw)$ \emph{and} in $\Im(\Mw)$ at all eight points, dominated by
the residual finite-difference truncation of the slope estimators;
the implementation thus transmits the magnetic effect through the
mass substitution consistently in the full complex frequency, with
no additional finite-radius magnetic input at $\mathcal{O}(bq)$.

\begin{table}[htbp]
\centering
\small
\begin{tabular}{ccc cc cc cc}
\toprule
$a/M$ & $\mu M$ & $m$ &
$\dfrac{\partial\Re(\Mw)}{\partial(\mu M)^{2}}$ &
$\dfrac{\partial\Im(\Mw)}{\partial(\mu M)^{2}}$ &
$\delta_{\rm pred}^{\Re}$ & $\delta_{\rm obs}^{\Re}$ &
$\delta_{\rm pred}^{\Im}$ & $\delta_{\rm obs}^{\Im}$ \\
\midrule
\multicolumn{9}{l}{\textit{$\mu M=0$, forward difference \eqref{slope.fd}}} \\
\midrule
$0.3$ & $0$   & $+1$ & $0.3854$ & $0.2122$ & $+0.000771$ & $+0.000770$ & $+0.000424$ & $+0.000422$ \\
$0.3$ & $0$   & $-1$ & $0.5007$ & $0.3209$ & $-0.001001$ & $-0.000998$ & $-0.000642$ & $-0.000637$ \\
$0.5$ & $0$   & $+1$ & $0.3352$ & $0.1681$ & $+0.000670$ & $+0.000669$ & $+0.000336$ & $+0.000334$ \\
$0.5$ & $0$   & $-1$ & $0.5321$ & $0.3511$ & $-0.001064$ & $-0.001061$ & $-0.000702$ & $-0.000697$ \\
\midrule
\multicolumn{9}{l}{\textit{$\mu M=0.3$, three-point one-sided on $(\mu M)^{2}\in\{0.01,0.04,0.09\}$}} \\
\midrule
$0.3$ & $0.3$ & $+1$ & $0.3942$ & $0.2473$ & $+0.000788$ & $+0.000786$ & $+0.000495$ & $+0.000498$ \\
$0.3$ & $0.3$ & $-1$ & $0.5172$ & $0.3669$ & $-0.001034$ & $-0.001030$ & $-0.000734$ & $-0.000736$ \\
$0.5$ & $0.3$ & $+1$ & $0.3434$ & $0.1982$ & $+0.000687$ & $+0.000685$ & $+0.000396$ & $+0.000399$ \\
$0.5$ & $0.3$ & $-1$ & $0.5530$ & $0.4006$ & $-0.001106$ & $-0.001101$ & $-0.000801$ & $-0.000804$ \\
\bottomrule
\end{tabular}
\caption{Internal consistency check of the implementation via the slope
relation \eqref{slope.prediction} at $qM=0.1$, $bM=10^{-2}$, $\ell=1$,
in both real and imaginary parts. The complex unmagnetized slope is
extracted from Tables~\ref{tab.Dolan2} and \ref{tab.Cp}: forward
difference \eqref{slope.fd} at $\mu M=0$, and the three-point one-sided
estimator on $(\mu M)^{2}\in\{0.01,0.04,0.09\}$ at $\mu M=0.3$. The
predicted shifts are
$\delta_{\rm pred}=[\partial(\Mw)/\partial(\mu M)^{2}]\cdot 2qbm M^{2}$;
the observed shifts are differences between magnetized and unmagnetized
roots at the same point. Predicted and observed values agree to
$\le 7\times 10^{-6}$ in $\Re(\Mw)$ and $\le 6\times 10^{-6}$ in
$\Im(\Mw)$ at every point, the residuals being dominated by the
finite-difference truncation of the slope estimators.}
\label{tab.slope}
\end{table}

At fixed $b$, the first-order shift is proportional to $q$.
Table~\ref{tab.linearity} compares $qM=0.05$ and $0.10$.
Using the displayed seven-decimal frequencies gives real- and
imaginary-shift ratios approximately $0.500$ and $0.499$.
Their small departures from one half are consistent with curvature
of the auxiliary Kerr spectrum and the rounding of displayed roots;
the printed data do not support six decimal places in these ratios.

\begin{table}[htbp]
\centering
\begin{tabular}{c cc cc}
\toprule
$qM$ & $\Re(\Mw)$ & $\Im(\Mw)$ &
$\delta\Re(\Mw)$ & $\delta\Im(\Mw)$ \\
\midrule
$0$ (ref.) & 0.3201265 & $-0.0966913$ & --- & --- \\
$0.05$ & 0.3205113 & $-0.0964807$ & $+0.0003848$ & $+0.0002106$ \\
$0.10$ & 0.3208962 & $-0.0962696$ & $+0.0007697$ & $+0.0004217$ \\
\bottomrule
\end{tabular}
\caption{Linearity check at $a=0.3M$, $\mu M=0$, $m=+1$, $\ell=1$,
$bM=10^{-2}$, with frequencies quoted to seven decimals. The
$q$-independent reference row is the unmagnetized root, identical for
both charges at this order. The differences in the last two columns are calculated from the
displayed frequencies. Their ratios are approximately $0.500$ and
$0.499$ in the real and imaginary parts; rounding these small
differences precludes a more precise ratio test from this table.}

\label{tab.linearity}
\end{table}

\subsection{Higher-order angular eigenvalue check}
\label{subsec.lambda.HO}

The continued-fraction implementation uses the leading approximation
\eqref{lambda.approx}, valid at small $c^{2}$, for the spheroidal
eigenvalue.
For the parameters of the magnetized survey $(qM\le 0.1$,
$bM\le 10^{-2}$, $|\Mw|\sim 0.3)$, $|c^{2}|\lesssim 0.1$, and the
next-order term in the Seidel expansion is parametrically of order
$10^{-4}$ in $\Lambda_{\ell m}$. To verify that this does not
contaminate the leading-order magnetic shift, we recompute the
magnetized spectrum at $bM=10^{-2}$ at eight representative
$(a,\mu M,m,\ell)$ points using the higher-order angular eigenvalue
$\Lambda_{\ell m}\simeq\ell(\ell+1)+f_{1}c^{2}+f_{2}c^{4}$
\cite{Seidel:1989bp,Berti:2005gp}, and compare with the leading-order
result. The coefficient $f_{2}$ is obtained here in closed form from
second-order perturbation theory with the exact matrix elements of
$x^{2}$ in the normalized associated-Legendre basis, giving
$f_{2}=-4/875\simeq -4.571\times 10^{-3}$ for $\ell=1$, $|m|=1$ and
$f_{2}\simeq -1.944\times 10^{-3}$ for $\ell=2$, $|m|=2$; this exact
second-order coefficient was checked against the direct spectral
solver introduced below (Table~\ref{tab.lambda.spec}); the two agree
to better than $10^{-9}$ in the resulting frequencies at every
sampled point. The comparison with the leading-order treatment is reported
in Table~\ref{tab.lambda.HO}, now for both $\Re(\Mw)$ and
$\Im(\Mw)$. At every sampled point the differences are at most a few
$\times 10^{-7}$, far below the four-decimal precision of the
spectral tables, confirming that the leading-order angular treatment
is adequate at the sampled points and the displayed survey precision.

\begin{table}[htbp]
\centering
\small
\setlength{\tabcolsep}{3.5pt}
\begin{tabular}{cccc|cc|cc|cc}
\toprule
$a/M$ & $\mu M$ & $m$ & $\ell$ &
$\Re(\Mw)_{f_{1}}$ & $\Im(\Mw)_{f_{1}}$ &
$\Re(\Mw)_{f_{1}+f_{2}}$ & $\Im(\Mw)_{f_{1}+f_{2}}$ &
$|\Delta\Re|$ & $|\Delta\Im|$ \\
\midrule
$0.3$ & $0$   & $+1$ & $1$ & $0.3208962$ & $-0.0962696$ & $0.3208962$ & $-0.0962696$ & 1.1e-08 & 2.6e-08 \\
$0.3$ & $0$   & $-1$ & $1$ & $0.2716363$ & $-0.0978646$ & $0.2716363$ & $-0.0978646$ & 3.9e-09 & 1.7e-08 \\
$0.3$ & $0.3$ & $+1$ & $1$ & $0.3559987$ & $-0.0756490$ & $0.3559987$ & $-0.0756490$ & 5.4e-09 & 6.4e-09 \\
$0.3$ & $0.3$ & $-1$ & $1$ & $0.3174058$ & $-0.0672010$ & $0.3174058$ & $-0.0672010$ & 4.1e-09 & 7.0e-10 \\
$0.5$ & $0$   & $+1$ & $1$ & $0.3454227$ & $-0.0940610$ & $0.3454226$ & $-0.0940607$ & 1.4e-07 & 2.5e-07 \\
$0.5$ & $0.3$ & $-1$ & $1$ & $0.3092913$ & $-0.0636906$ & $0.3092913$ & $-0.0636906$ & 2.7e-08 & 1.7e-09 \\
$0.3$ & $0.3$ & $+2$ & $2$ & $0.5614499$ & $-0.0880774$ & $0.5614498$ & $-0.0880774$ & 2.0e-08 & 2.5e-08 \\
$0.3$ & $0.3$ & $-2$ & $2$ & $0.4760987$ & $-0.0845278$ & $0.4760987$ & $-0.0845278$ & 6.0e-09 & 1.3e-08 \\
\bottomrule
\end{tabular}
\caption{Higher-order angular eigenvalue check at $qM=0.1$,
$bM=10^{-2}$, at truncation depth $N=120$. The columns labelled
$f_{1}$ use the leading Seidel approximation
$\Lambda_{\ell m}\simeq\ell(\ell+1)+f_{1}c^{2}$; the columns labelled
$f_{1}+f_{2}$ add the exact second-order coefficient $f_{2}c^{4}$
derived in the text ($f_{2}=-4/875$ for $\ell=1$, $|m|=1$). Both
$\Re(\Mw)$ and $\Im(\Mw)$ are reported to seven decimals; the last
two columns give the absolute differences, at most a few
$\times 10^{-7}$ at every sampled point, justifying the use of the
leading-order angular eigenvalue throughout the magnetized survey.}
\label{tab.lambda.HO}
\end{table}

Both columns of Table~\ref{tab.lambda.HO} rest on truncations of the
\emph{same} small-$c^{2}$ Seidel expansion, so their agreement bounds
the $f_{2}c^{4}$ term but not, by itself, the full expansion error.
To close this gap we also computed the angular eigenvalue
\emph{non-perturbatively}, by solving the spheroidal eigenvalue
problem directly at the complex spheroidicity
$c^{2}=a^{2}(\omega^{2}-\mueff^{2})$ of each root. Expanding
$S_{\ell m}$ in the basis of normalized associated Legendre functions
$P_{\ell'}^{m}$, the angular equation \eqref{spheroidal.eq} becomes the
matrix eigenvalue problem
\be
\Bigl[\,{\rm diag}\bigl(\ell'(\ell'+1)\bigr)\;-\;c^{2}\,X^{2}\Bigr]
\vec S \;=\;\Lambda_{\ell m}\,\vec S ,
\qquad
X_{\ell'\!,\ell'+1}=X_{\ell'+1,\ell'}
=\sqrt{\frac{(\ell'+1)^{2}-m^{2}}{(2\ell'+1)(2\ell'+3)}},
\label{spectral.method}
\ee
where $X$ is the exact tridiagonal representation of $\cos\theta$ in
that basis; we truncate at $45$ basis functions, far beyond
convergence at $|c^{2}|\lesssim 0.1$. The eigenvalue closest to the
Seidel estimate is selected and the continued-fraction root is then
re-solved with $\Lambda_{\ell m}$ re-evaluated by
\eqref{spectral.method} at every Newton iterate. Table~\ref{tab.lambda.spec}
compares the resulting frequencies with the production (leading
Seidel) values, in \emph{both} real and imaginary parts. The
differences do not exceed $1.4\times 10^{-7}$ in $\Re(\Mw)$ and
$2.5\times 10^{-7}$ in $\Im(\Mw)$, the largest occurring at $a=0.5M$
where $|c^{2}|$ is largest. The leading-order angular treatment is
therefore adequate for the displayed precision at the sampled points, as supported by the sampled expansion-independent comparisons.

\begin{table}[htbp]
\centering
\begin{tabular}{ccc cc cc}
\toprule
$a/M$ & $\mu M$ & $m$ &
$\Re(\Mw)_{\rm spec}$ & $\Im(\Mw)_{\rm spec}$ &
$|\Delta\Re(\Mw)|$ & $|\Delta\Im(\Mw)|$ \\
\midrule
$0.3$ & $0$   & $+1$ & $0.32089621$ & $-0.09626961$ & $1.1\times10^{-8}$ & $2.6\times10^{-8}$ \\
$0.3$ & $0$   & $-1$ & $0.27163626$ & $-0.09786462$ & $3.9\times10^{-9}$ & $1.7\times10^{-8}$ \\
$0.3$ & $0.3$ & $+1$ & $0.35599868$ & $-0.07564896$ & $5.4\times10^{-9}$ & $6.4\times10^{-9}$ \\
$0.3$ & $0.3$ & $-1$ & $0.31740578$ & $-0.06720102$ & $4.1\times10^{-9}$ & $7.0\times10^{-10}$ \\
$0.5$ & $0$   & $+1$ & $0.34542259$ & $-0.09406074$ & $1.4\times10^{-7}$ & $2.5\times10^{-7}$ \\
$0.5$ & $0$   & $-1$ & $0.26051166$ & $-0.09720222$ & $2.2\times10^{-8}$ & $1.1\times10^{-7}$ \\
$0.5$ & $0.3$ & $+1$ & $0.37596940$ & $-0.07762282$ & $3.4\times10^{-8}$ & $8.5\times10^{-8}$ \\
$0.5$ & $0.3$ & $-1$ & $0.30929128$ & $-0.06369056$ & $2.7\times10^{-8}$ & $1.7\times10^{-9}$ \\
\bottomrule
\end{tabular}
\caption{Direct (non-perturbative) angular-eigenvalue check at
$\ell=1$, $qM=0.1$, $bM=10^{-2}$: QNM frequencies recomputed with the
spectral spheroidal eigenvalue \eqref{spectral.method} evaluated at
complex $c^{2}$ inside the root search, and the absolute differences
$|\Delta\Mw|$ from the production values obtained with the leading
Seidel approximation \eqref{lambda.approx}. Both real and imaginary
parts are compared; all differences are below
$2.5\times 10^{-7}$, well under the numerical error floor
$\delta\omega_{\rm num}$ of Sec.~\ref{subsec.precision}.}
\label{tab.lambda.spec}
\end{table}

\subsection{Extension to \texorpdfstring{$\ell=2$}{ell=2}}
\label{subsec.l2}

The structural result that the weakly magnetized charged-scalar problem
can be cast in the Kerr massive-scalar form under
$\mu^{2}\to\mueff^{2}=\mu^{2}+2qbm$ is independent of $\ell$. To verify
that the framework operates without modification at higher angular
momentum, we extend the calculation to $\ell=2$, $m=\pm 2$ at
$a/M=0.3$, $\mu M=0.3$, $qM=0.1$, for $bM\in\{0,10^{-3},10^{-2}\}$. The
results are reported in Table~\ref{tab.l2}.

The same opposite-sign pattern in $m$ persists at $\ell=2$: the
co-rotating mode ($m=+2$) shifts upward in $\Re(\Mw)$ as $b$ increases,
and the counter-rotating mode ($m=-2$) shifts downward. The shift relation \eqref{slope.prediction} uses the derivative at
the same scalar mass as the magnetized mode. At
$(a/M,\ell,m,\mu M)=(0.3,2,2,0.3)$, the quadratic interpolant through
the three nonzero-mass entries of Table~\ref{tab.Dolan3} gives
\be
\frac{\partial(M\omega)}{\partial(\mu M)^2}
\simeq 0.26295+0.08490\,i.
\label{l2.correct.slope}
\ee
With $2(qM)(bM)m=0.004$, this predicts
$\delta(M\omega)\simeq1.052\times10^{-3}
+3.396\times10^{-4}i$. The real shift agrees with the value
$0.5614499-0.560397\simeq1.053\times10^{-3}$ from
Tables~\ref{tab.lambda.HO} and \ref{tab.Dolan3}, to the accuracy
appropriate to the finite-difference derivative and displayed roots.
The $\ell=2$ shifts are controlled by their own Kerr mass slopes;
they are not obtained by simply scaling the $\ell=1$ shifts by
$|m|$.

\begin{table}[htbp]
\centering
\begin{tabular}{lllcccc}
\toprule
$\ell$ & $m$ & $bM$ &
$\Re(\Mw)$ & $\Im(\Mw)$ & $\tau/M$ & $Q$ \\
\midrule
$2$ & $+2$ & $0$        & 0.5604 & -0.0884 & 11.310 & 3.169 \\
$2$ & $+2$ & $10^{-3}$  & 0.5605 & -0.0884 & 11.314 & 3.171 \\
$2$ & $+2$ & $10^{-2}$  & 0.5614 & -0.0881 & 11.354 & 3.187 \\
\midrule
$2$ & $-2$ & $0$        & 0.4776 & -0.0840 & 11.909 & 2.844 \\
$2$ & $-2$ & $10^{-3}$  & 0.4774 & -0.0840 & 11.901 & 2.841 \\
$2$ & $-2$ & $10^{-2}$  & 0.4761 & -0.0845 & 11.830 & 2.816 \\
\bottomrule
\end{tabular}
\caption{Fundamental auxiliary effective-Kerr QNM frequencies at $\ell=2$, $n=0$,
$a=0.3M$, $\mu M=0.3$, $qM=0.1$, in the reference gauge.
The same continued-fraction implementation that produced
Table~\ref{tab.QNM.MK} for $\ell=1$ is applied without modification at
$\ell=2$. The opposite-sign pattern between the two $|m|=2$ sectors
persists, with magnitudes set by the corresponding $\ell=2$ unmagnetized
slopes via \eqref{slope.prediction}. Conventions as in
Table~\ref{tab.QNM.MK}: $\Phi\sim e^{-i\omega t}$, $\Im(\Mw)<0$ for a
damped mode, and $Q$ is the quality factor.}
\label{tab.l2}
\end{table}

\section{Conclusions}
\label{sec.conclusion}

We have formulated an auxiliary effective Kerr spectral problem
from the local first-order charged-scalar equation in the neutral-seed
Melvin--Kerr family. In the reference gauge, $A_\mu^{(b)}=b g_{\mu\phi}$
makes the retained interaction a constant $2qbm$ on each azimuthal
mode. Both separated equations therefore have the massive-scalar
Kerr form with $\mueff^2=\mu^2+2qbm$. The contraction of the
first-order potential with the horizon generator vanishes in this
gauge, and the invariant horizon phase reduces to $\omega-m\Omega_H$.

With Kerr boundary conditions imposed on the extended equation,
the auxiliary spectrum is the Kerr spectrum evaluated at this
shifted mass-squared parameter. Equation~\eqref{auxiliary.response}
expresses its linear response through the complex Kerr mass
derivative. The fundamental $\ell=1,2$ examples illustrate this
response, its dependence on $m$, and the smooth continuation of the
selected damped branch through $\mueff^2=0$. The slope tests are
internal checks of an algebraic identity, and the real-frequency
potential plots are qualitative diagnostics. Neither provides
independent evidence for the full magnetized spectrum.

The numerical checks support the reported precision of the auxiliary
roots in the surveyed range. They do not measure the effect of
omitted metric and quadratic electromagnetic terms, or the effect
of the exact exterior boundary conditions. Local power counting
alone yields no percentage accuracy for a global resonance. The
negative-effective-mass continuation tests particular QNM branches
and does not establish full linear stability.

The remaining physical task is to specify the outer problem for
the rotating magnetized geometry and relate it to the inner
solution. This could be done by controlled matching, a direct global
wave calculation, or an independently specified exterior completion.
Moving the same auxiliary outgoing solution to a finite matching
radius would only restate the auxiliary problem. Inclusion of
$\mathcal O(b^2)$ metric corrections would additionally require the
corresponding electromagnetic terms and, in general, a treatment
beyond the separability used here. These steps would determine
whether any part of the auxiliary response describes physical
magnetized resonances.

\section*{Acknowledgments}

Support for this work was provided by Direktorat Penelitian
dan Pengabdian kepada Masyarakat, Direktorat
Jenderal Riset dan Pengembangan, Kementerian
Pendidikan Tinggi, Sains dan Teknologi Republik
Indonesia in the year 2026 with contract number
\path{1978/LL4/PG/2026}; \path{III/LPPM/2026-05/183-PE} and
\path{131/C3/DT.05.00/PL-MULTITAHUN LANJUTAN/2026}.

\section*{Declaration of generative AI and AI-assisted technologies in the manuscript preparation process}
During the preparation of this work the author used Claude (Anthropic)
and ChatGPT (OpenAI) for language editing, restructuring the manuscript, and assistance with algebraic checks and numerical implementations. The author reviewed and edited the output as needed and takes full responsibility for the content of the published article. 

\appendix

\section{The first-order gauge potential and the horizon
electrostatic potential}
\label{app.horizon}

This appendix demonstrates explicitly, at the level of the gauge
potential itself, why the horizon boundary condition of
Sec.~\ref{sec.QNM} carries no $\mathcal{O}(bq)$ magnetic correction.
We display the $\mathcal{O}(b)$ potential in closed form, evaluate the
horizon Killing potential $\chi^{\mu}A_{\mu}|_{r_{+}}$ with
$\chi=\partial_{t}+\Omega_{H}\partial_{\phi}$ the horizon generator,
and trace the physical origin of its vanishing.

\subsection*{A.1\quad The first-order potential in the
reference gauge}

Since the Kerr seed of the Harrison transformation is electrically
neutral, the exact Melvin--Kerr potential is odd in $b$ at leading
order, and its linear-in-$b$ part is a stationary, axisymmetric,
source-free test Maxwell field on the exact Kerr background. In the
uniform-field/monopole sector relevant to the weak-$b$ Harrison
expansion, the natural ansatz is a linear combination of Wald's
uniform-field solution \cite{Wald:1974np}, built from the Killing
one-forms $\eta_{\mu}=g_{\mu\nu}(\partial_{t})^{\nu}$ and
$\psi_{\mu}=g_{\mu\nu}(\partial_{\phi})^{\nu}$, the Kerr--Newman
charge mode, and the residual constant $dt$. Matching this ansatz
against the separated coefficients
\eqref{fmu.PLB}--\eqref{f0.PLB} fixes its coefficients uniquely
(equivalently, the result below can be verified directly by expanding
the exact Harrison-transformed potential to $\mathcal{O}(b)$), with a
remarkably compact outcome: in the reference gauge
\eqref{gauge.shift}, the entire $\mathcal{O}(b)$ potential---denoted
$A^{(b)}_{\mu}\equiv b\,{\cal A}^{(1)}_{\mu}$ in the expansion
notation of Sec.~\ref{sec.MK}---is the metric-dual of the axial
Killing vector,
\be
A^{(b)}_{\mu}\,dx^{\mu}\;=\;b\,\psi_{\mu}\,dx^{\mu}
\;=\;b\left[\,g_{t\phi}\,dt+g_{\phi\phi}\,d\phi\,\right],
\label{app.A.psi}
\ee
i.e., explicitly, with $\Sigma=r^{2}+a^{2}\cos^{2}\theta$,
\be
A^{(b)}_{t}=\frac{2abMr\sin^{2}\theta}{\Sigma},
\qquad
A^{(b)}_{\phi}
=-\,b\left[(r^{2}+a^{2})+\frac{2a^{2}Mr\sin^{2}\theta}{\Sigma}\right]
\sin^{2}\theta .
\label{app.A.components}
\ee
One checks directly that $A^{(b)}_{t}\to 0$ as $r\to\infty$, as
required by the gauge condition, and that adding the constant
$+\,4M^{2}b\,a^{-1}\,dt$ of \eqref{gauge.shift} reproduces the
$\mathcal{O}(b)$ part of the Harrison-gauge potential of
Ref.~\cite{Siahaan:2025plb} (its coefficient functions $k_{0}$ and
$l_{0}$ at $b^{0}$ are $r(a^{2}+2Mr)$ and $-r(r^{3}+a^{2}r+2a^{2}M)$,
as \eqref{app.A.components} gives), including the asymptotic value
\eqref{At.Harrison.inf}.

The verification against
\eqref{fmu.PLB}--\eqref{f0.PLB} is sharpest in contravariant form,
where \eqref{app.A.psi} collapses to
\be
A^{(b)\,t}=0,
\qquad
A^{(b)\,\phi}=b ,
\label{app.A.contravariant}
\ee
a \emph{constant} vector field along $\partial_{\phi}$. Inserting
\eqref{app.A.contravariant} into the $\mathcal{O}(bq)$ interaction
terms of \eqref{KG.eq},
$-iq\,(2A^{\mu}\partial_{\mu}+\nabla_{\mu}A^{\mu})\Phi$, and using the
mode ansatz \eqref{ansatz}: the divergence vanishes exactly,
$\nabla_{\mu}A^{(b)\mu}=b\,\nabla_{\mu}\psi^{\mu}=0$, by Killing's
equation, while $2A^{(b)\mu}\partial_{\mu}\Phi=2b\,\partial_{\phi}\Phi
=2ibm\,\Phi$, so the full interaction is the constant multiple
$+2qbm\,\Phi$ and the retained equation is
$(\Box_{\rm Kerr}+\mu^{2}+2qbm)\Phi=0$. After the overall $-\Sigma$
rescaling of \eqref{KG.general.structure} this is precisely
$f_{m}m=-2qbm\,\Sigma$ alongside $f_{\mu^{2}}\mu^{2}=-\Sigma\mu^{2}$,
with no magnetic contribution to any other coefficient---in
particular $A^{(b)\,t}=0$ explains structurally why $f_{\omega}$
carries no magnetic term. (In the flat cylindrical limit,
$A_{\phi}=-b\varrho_{\rm cyl}^{2}$ gives the azimuthal operator
$+(m+qb\varrho_{\rm cyl}^{2})^{2}/\varrho_{\rm cyl}^{2}$, whose
cross term is the same $+2qbm$: the standard Zeeman-type coupling
of a charge $q$ to a uniform field $B_{0}=2b$ in the convention
$D_{\mu}=\nabla_{\mu}-iqA_{\mu}$.) The mass substitution
$\mu^{2}\to\mueff^{2}$ of Sec.~\ref{sec.KG}, including the absence of
any finite-radius magnetic residue, is thereby re-derived in two lines
from the geometry of the potential.

\subsection*{A.2\quad Vanishing of the horizon Killing potential}

For a mode $\Phi\sim e^{i(m\phi-\omega t)}$, the phase gradient along
the horizon generator is
\be
\chi^{\mu}\bigl(\partial_{\mu}-iqA_{\mu}\bigr)\Phi
=-\,i\bigl[\,\omega-m\Omega_{H}+q\,\chi^{\mu}A_{\mu}\,\bigr]\Phi ,
\label{app.chi.phase}
\ee
so the near-horizon indicial exponent is controlled by the
combination $\omega-m\Omega_{H}+q\,\chi^{\mu}A_{\mu}|_{r_{+}}$, which
is invariant under the residual gauge freedom: a constant shift
$A_{t}\to A_{t}+C$ moves $\chi^{\mu}A_{\mu}|_{r_{+}}$ by $+C$ and the
frequency by $\omega\to\omega-qC$
[Eq.~\eqref{gauge.dictionary}], leaving \eqref{app.chi.phase}
unchanged. In the reference gauge, \eqref{app.A.psi}
gives
\be
\chi^{\mu}A^{(b)}_{\mu}\big|_{r_{+}}
=b\,(\chi\cdot\psi)\big|_{r_{+}}
=0 ,
\label{app.chiA.zero}
\ee
\emph{identically} at $\mathcal{O}(b)$, for a purely geometric reason:
on the Killing horizon, $\chi$ is the null normal while the axial
Killing vector $\partial_{\phi}$ (whose metric-dual is $\psi$) is
tangent, and a null normal is orthogonal to every horizon-tangent
vector, so $g(\chi,\partial_{\phi})|_{\mathcal{H}}=
\chi^{\mu}\psi_{\mu}|_{\mathcal{H}}=0$. (The same conclusion
follows by direct evaluation of
\eqref{app.A.components} at $r=r_{+}$ using
$a^{2}=2Mr_{+}-r_{+}^{2}$; the result is $\theta$-independent, as
required by the constancy of the horizon electrostatic potential.) The gauge-invariant near-horizon
combination therefore reduces, at $\mathcal{O}(bq)$ and with the
frequency of the reference gauge, exactly to the Kerr
form $\omega-m\Omega_{H}$: the horizon exponent $\sigma$ of
Sec.~\ref{sec.QNM} is the unmodified Kerr one as a demonstrated,
gauge-invariant statement, not merely as a feature of the separated
coefficient list.

\subsection*{A.3\quad Induced Wald charge versus residual gauge}

It is instructive to ask whether the cancellation
\eqref{app.chiA.zero} is ``physical'' or ``gauge''. The answer is an
exact interplay of both, made transparent by decomposing
\eqref{app.A.psi} in the basis introduced above:
\be
b\,\psi_{\mu}
=\underbrace{b\,\bigl(\psi_{\mu}+2a\,\eta_{\mu}\bigr)}_{\text{Wald
uniform-field mode}}
\;+\;\underbrace{Q_{\rm ind}\,\frac{r}{\Sigma}
\bigl(dt-a\sin^{2}\theta\,d\phi\bigr)_{\mu}}_{\text{induced charge
mode}}
\;-\;\underbrace{2ab\;(dt)_{\mu}}_{\text{constant}},
\qquad
Q_{\rm ind}=4abM ,
\label{app.decomposition}
\ee
where the charge mode is the Kerr--Newman potential
$(Qr/\Sigma)(dt-a\sin^{2}\theta\,d\phi)$ in the present signature.
Identifying the weak-field strength $B_{0}=2b$ from the contravariant
component $A^{(b)\phi}=b=B_{0}/2$ (equivalently from the large-radius
flux function $A^{(b)}_{\phi}\to -(B_{0}/2)\,r^{2}\sin^{2}\theta$ of
the $(+---)$ signature), the induced charge is
$Q_{\rm ind}=2B_{0}J$ with $J=aM$: the Harrison transformation of the
neutral Kerr seed endows the $\mathcal{O}(b)$ solution with precisely
the Wald equilibrium charge \cite{Wald:1974np} (for the physical
charge of magnetized black holes see also \cite{Gibbons:2013yq}). Evaluating the horizon
Killing potential of each piece: the Wald mode contributes zero, since the Killing vectors
$\partial_{t}$ and $\partial_{\phi}$ (with metric-duals $\eta$ and
$\psi$) are tangent to the horizon and hence orthogonal to $\chi$
there; the charge mode contributes the standard
Kerr--Newman horizon potential
$Q_{\rm ind}\,r_{+}/(r_{+}^{2}+a^{2})=Q_{\rm ind}/(2M)=2ab$; and the
constant contributes $-2ab$. The sum vanishes:
\be
\chi^{\mu}A^{(b)}_{\mu}\big|_{r_{+}}
=0\;+\;2ab\;-\;2ab\;=\;0 .
\label{app.split}
\ee
Thus the vanishing of the horizon potential in the reference gauge combines the genuinely physical potential of the induced
Wald charge with the residual gauge constant fixed by the condition
$A_{t}(\infty)=0$; what is gauge-independent is the full combination
\eqref{app.chi.phase}.

\subsection*{A.4\quad Cross-check in the Harrison gauge}

In the unshifted Harrison gauge, $A^{\rm (Harrison)}_{t}
=A^{(b)}_{t}+4M^{2}b/a$, so
\be
\chi^{\mu}A^{\rm (Harrison)}_{\mu}\big|_{r_{+}}=\frac{4M^{2}b}{a},
\label{app.chiA.Harrison}
\ee
while the frequency dictionary of Sec.~\ref{sec.MK} gives
$\omega^{\rm (Harrison)}=\omega-4qM^{2}b/a$; the invariant
combination \eqref{app.chi.phase} is identical in the two gauges, as
it must be. This is consistent with the near-horizon radial momentum
obtained in the Harrison gauge in Ref.~\cite{Siahaan:2025plb}, which
takes the form $k_{+}^{2}=(\omega-\omega_{c})^{2}
+(8bM^{2}q/a)(\omega-\omega_{c})$ with $\omega_{c}=m\Omega_{H}$:
to the order retained this is
$[\,(\omega-\omega_{c})+4bqM^{2}/a\,]^{2}+\mathcal{O}(b^{2})$, i.e.\
precisely the square of the invariant
$\omega-m\Omega_{H}+q\,\chi^{\mu}A_{\mu}|_{r_{+}}$ evaluated with the
Harrison-gauge quantities. Re-expressed through the frequency of the
reference gauge, the $\mathcal{O}(bq)$ shift disappears,
in agreement with \eqref{app.chiA.zero}.

\end{document}